\documentclass{article}
\usepackage{graphicx}
\graphicspath{ {./images/} }
\usepackage[utf8]{inputenc}
\usepackage{caption}
\usepackage{subcaption}
\usepackage{enumerate}
\usepackage[square,sort,comma,numbers]{natbib}
\usepackage{listings}
\usepackage{caption}
\usepackage{array}
\usepackage{mathtools}
\usepackage{graphicx}
\usepackage{algorithm}
\usepackage{algpseudocode}
\usepackage{multirow}
\usepackage{array}
\usepackage{color}
\usepackage{subcaption}
\usepackage{multirow}
\usepackage{tikz}
\title{LWA/LWIP document}
\author{Shahwaiz Afaqui}
\date{October 2018}
\usepackage[symbol]{footmisc}
\def\correspondingauthor{\footnote{Corresponding author: M. Shahwaiz Afaqui, Email: mafaqui@uoc.edu}}
\title{Implementation of the 3GPP LTE-WLAN Interworking Protocols in NS-3}
\usepackage{authblk}
\author[1]{ M. Shahwaiz Afaqui\correspondingauthor{}}
\author[2]{Cristina Cano}
\author[3]{Vincent Kotzsch} 
\author[4]{Clemens Felber}
\author[5]{Walter Nitzold}
\affil[1,2]{Wireless Network Group (WiNe), Universitat Oberta de Catalunya(UOC), Barcelona, Spain.}
\affil[3,4,5]{National Instruments, 
Dresden, Germany}
\date{\today}
\newcommand{\chg}[1]{{\color{black} #1}}
\begin{document}
\maketitle

\begin{abstract}
The next generation wireless standard, called Fifth Generation (5G), is being designed to encompass Heterogeneous Networks (HetNets) architectures consisting of a single holistic network with Multiple Radio Access Technologies (Multi-RAT). Multiple connectivity protocols and spectrum would be managed from a common core (management system) handling both: \emph{i)} traditional macro-cellular systems (such as LTE), that can provide long-range, outdoor coverage, as well as \emph{ii)} low-power wireless systems with high capacity (such as Wi-Fi), that can be deployed to cater indoor traffic needs. 5G HetNets are expected to achieve ubiquitous connectivity that would guarantee Quality of Service (QoS), Quality of Experience (QoE) along with efficient use of spectrum and energy at low cost. Tightly coupled LTE—Wi-Fi networks have emerged as one of the promising solutions in the 5G era to boost network capacity and improve end user's quality of experience. LTE/Wi-Fi Link Aggregation (LWA) and LTE WLAN Radio Level Integration with IPSec Tunnel (LWIP) are two approaches put forward by the 3rd Generation Partnership Project (3GPP) to enable flexible, general, and scalable LTE-WLAN inter-working. These techniques enable operator-controlled access of licensed and unlicensed spectrum and allow transparent access of operator's evolved core. The most important aspect of these techniques is that they could be enabled with straightforward software upgrades and can utilize the already existing Wi-Fi networks. This article presents and motivates the design details of LWA and LWIP protocols. We also present the first NS-3 LWA and LWIP implementations over Network Simulator 3 (NS-3). In particular, this work focuses on the adaptation and concurrent usage of different NS-3 modules and protocols of different technologies to enable the support of these interworking schemes. 

\end{abstract}

\section{Introduction}
The dramatic increase in the number of smart phones, tablets, wearable, and other smart mobile devices has resulted in tremendous growth in traffic demands for current and future wireless networks. According to Cisco Visual Networking Index (VNI) \cite{cisco}, global mobile data traffic will reach 48.3 EB per month by 2021 with a growth of 46 percent between 2016 and 2021.

Thus, future wireless networks are expected to manage and satisfy the requirements of billions of connected
devices. Due to the near-Shannon’s capacity limit achieved by the current cellular wireless technologies (such as LTE Advanced and LTE Pro) the foreseen growth in traffic can only be sustained by: a) densification of wireless networks, b) by deploying Heterogeneous Network (HetNet) architectures consisting of a single holistic network that include network deployments of nodes
which support multiple Radio Access Technology (RAT) joint operations.

Therefore, higher capacity and data rate along with improved spectral efficiency and reduced power consumption
requirements are anticipated to be fulfilled by dense HetNet deployments, where IEEE 802.11-based
Wi-Fi networks and LTE-based cellular networks are envisioned to be the main technologies whose joint
operations can provide indoor as well as outdoor coverage. Currently, LTE operates efficiently over the licensed
spectrum from 700MHz to 2.6GHz, while Wi-Fi operates on Industrial, Scientific and Medical (ISM) 2.4 GHz
and National Information Infrastructure (U-NII) 5 GHz band.

The IEEE 802.11-based Wireless Local Area Networks (WLANs) (i.e. Wi-Fi networks), are the most successful
indoor wireless solutions and have evolved as a key technology adopted in unlicensed bands to cover
medium to large-scale enterprise, public area hot-spots and apartment complexes, among others. Due to the continuous
technological innovations and their ability to provide increased mobility, flexibility, ease of use along with
reduced cost of setting up, Wi-Fi continues to remain the first choice of connectivity for numerous use cases.
Motivated by the scarcity and cost of licensed spectrum, the future small cells HetNets based on cellular
technologies are expected to share the same spectrum with macro-cells, that will result in severe co-channel
interference. Motivated to achieve the required future capacity growth and to reduce the impact of co-channel
interference, standardization efforts are underway by the 3GPP to enable cellular mobile networks operate over the unlicensed 2.4 and 5GHz spectrum which is currently being used by Wi-Fi, Zigbee and
other communication systems.

To date, there are five different unlicensed LTE approaches:  1) LTE-WLAN Radio Aggregation (LWA) \cite{LWA-LWIP}, 2) LTE-WLAN Radio Level Integration (LWIP)\cite{LWA-LWIP}, 3) Licensed Assisted Access (LAA)\cite{LAA}, 4) LTE-Unlicensed (LTE-U)\cite{LTE-U}, and 5) MuLTEFire\cite{multifire}. Each of these alternative methods proposed for LTE to offload data traffic over unlicensed spectrum aim to opportunistically meet the future growth in cellular traffic. However, no analysis is found in the
state-of-the-art which compares and ranks these techniques according to their feasibility of implementation and performance. 

In this work, we present the implementation details of LWA and LWIP protocols over Network Simulator 3 (NS-3)\cite{NS-3} to enable filling the aforementioned gap. NS-3 is a discrete-event network simulator for internet systems and is developed in C++ language. It provides realistic models to mimic the behavior of packet data networks. NS-3 provides a great compromise, by combining the ability to run real applications and network protocol codes, with the flexibly, as well as the ability to simulate in a controlled network environment and ease of reproducibility. NS-3 also provides support for several models and protocols such as Wi-Fi, WiMAX, LTE, Point-to-Point and so on.
The details of different NS-3 modules used and enhanced to develop the aforementioned techniques are provided. Our design is based on specifications of LWA and LWIP set in Release 13 of 3GPP.   

\paragraph{Outline}
The remainder of this article is organized as follows. In Sections \ref{Overview of LWA architecture} and \ref{Overview of LWIP architecture}, a detailed overview of LWA and LWIP mechanisms is presented, respectively. Section \ref{Background} enlists and explains the building blocks that are used within the design of these interworking techniques in NS-3. Section \ref{Customization to create LWA model in NS-3} and \ref{Customization to create LWIP model in NS-3} describe the implementation details of LWA and LWIP. Validation results for the two techniques are presented in Section \ref{results}. Additional results of LWA technique implemented over two separate instances of NS-3 are highlighted in Section \ref{Results of LWA and LWIP for two instance of NS-3 simulation}. Finally, Section~\ref{conclusions} presents conclusions of this article.
\section{Overview of LWA architecture}
\label{Overview of LWA architecture}
LTE-WLAN Aggregation (LWA), was first introduced in 3GPP release 13 and combines Wi-Fi unlicensed bandwidth with the licensed LTE bandwidth, taking advantage of the high indoor availability of Wi-Fi networks. It efficiently integrates LTE and WLAN at the Packet Data Convergence Protocol (PDCP) layer of LTE. LWA aims for optimal operation of licensed/unlicensed bands by allowing downlink traffic to be carried by both LTE and WLAN. In order to avoid the asymmetric contention problem of Wi-Fi, the uplink traffic is only carried through LTE. Wi-Fi APs, which act as secondary access for user data, are connected to LWA base stations and thus can leverage LTE core network functionalities without a dedicated gateway.
\begin{figure}
\centering
\begin{subfigure}[b]{.5\textwidth}
  {\raisebox{12mm}{
  \includegraphics[width=1\linewidth]{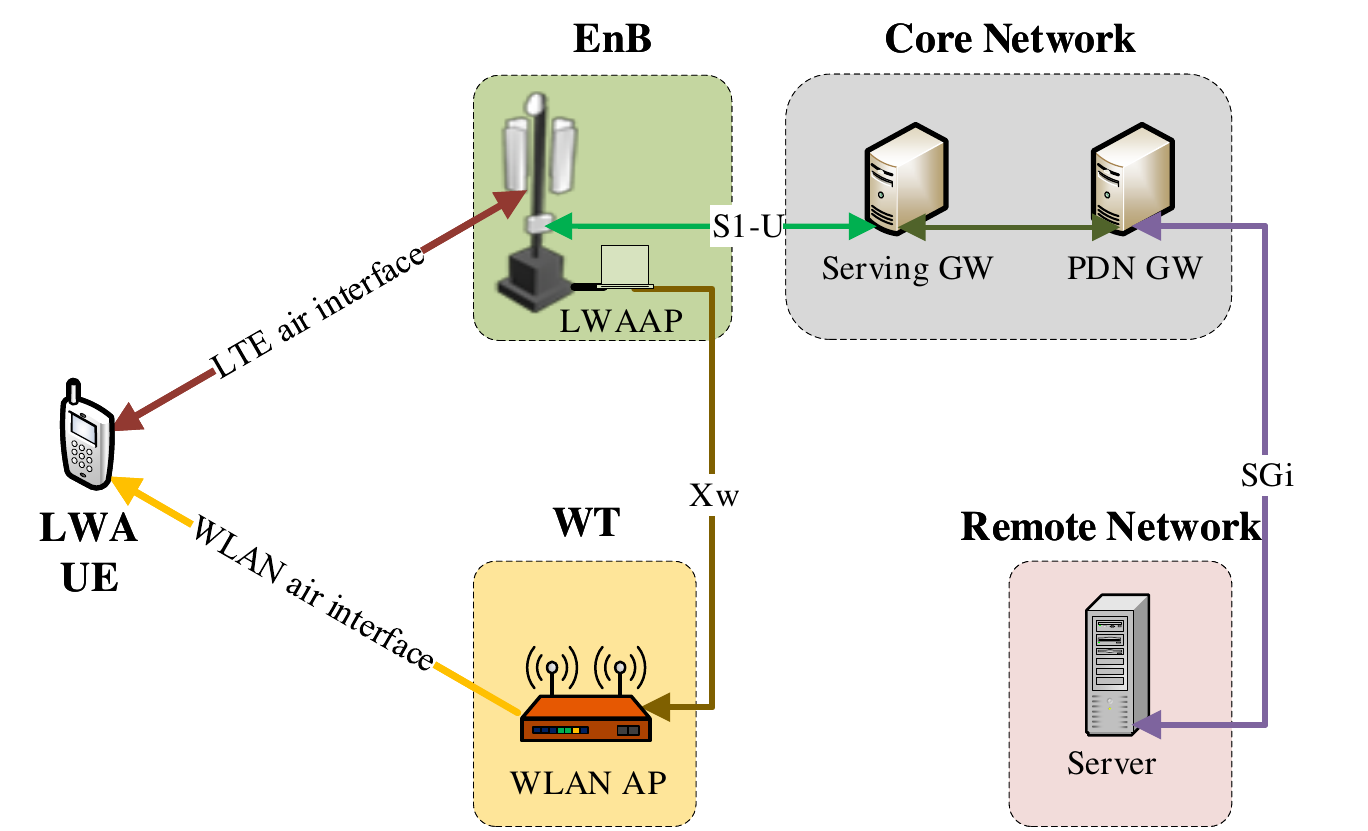}}}
  \caption{Network Architecture}
  \label{fig1:sub1}
\end{subfigure}%
\begin{subfigure}[b]{.5\textwidth}
  \centering
  \includegraphics[width=1\linewidth]{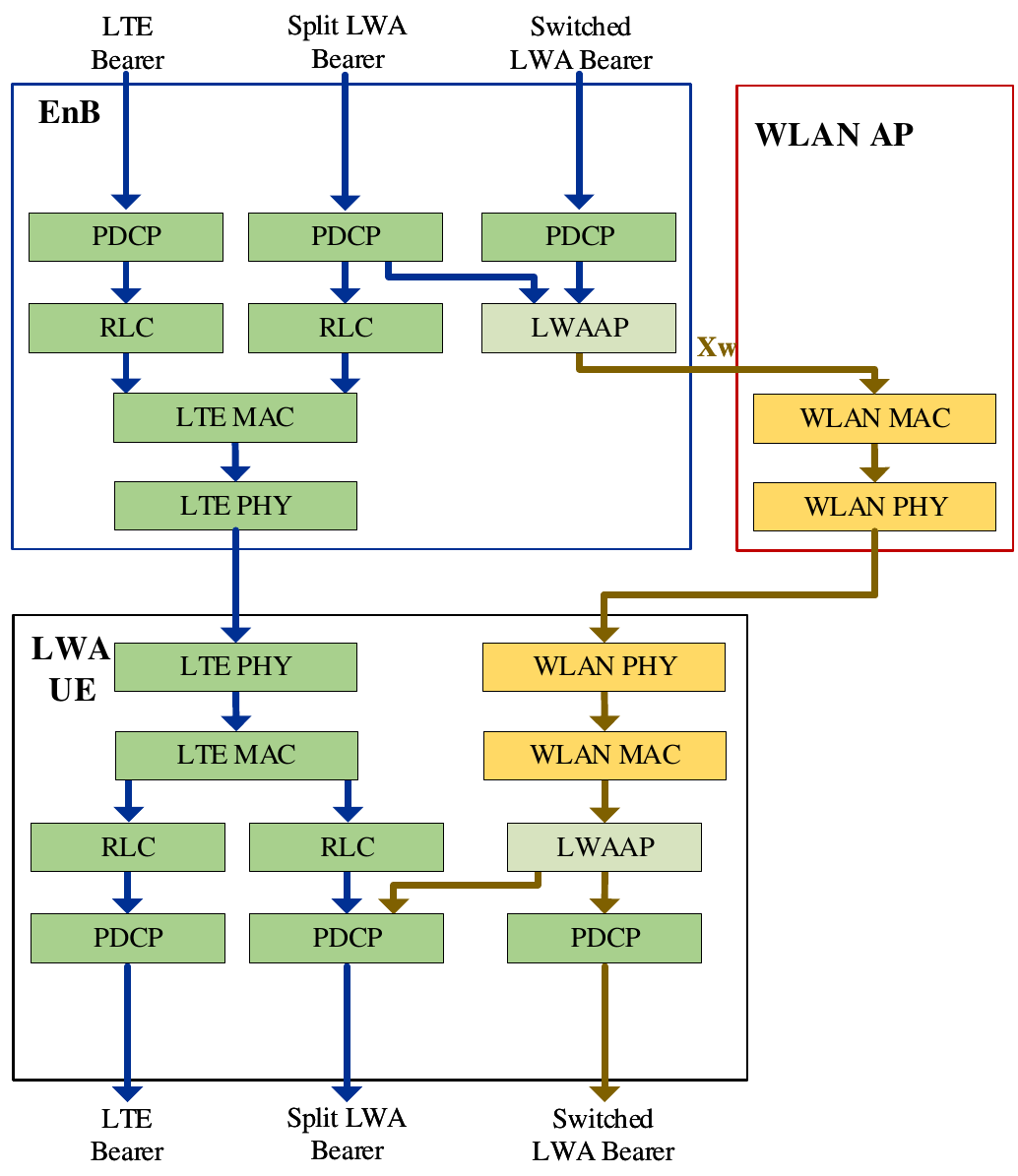}
  \caption{Protocol architecture defined by 3GPP}
  \label{fig1:sub2}
\end{subfigure}
\caption{Overview of LWA}
\label{fig1}
\end{figure}

LWA architecture consists of Evolved Node B (eNB), LWA-aware Wi-Fi Access Point (AP), LWA-aware Wi-Fi station and User Equipment (UE). Release 13 of 3GPP LTE defines two development scenarios of LWA, namely collocated and non-collocated, through which the eNB and WLAN entities are connected. In the collocated scenario, the Wi-Fi AP is collocated and connected through an internal backhaul connection to the eNB. This deployment option is more appropriate for small cells. For the non-collocated deployment as shown in \chg{Figure \ref{fig1:sub1}}, an optional standardized interface, called $Xw$, is used to connect the WLAN AP to the eNB through a WLAN Termination (WT) logical node (which can be a Wi-Fi AP or a Wi-Fi controller). This interface supports control (called $Xw$-$C$) and user (data) plane (called $Xw$-$U$). Apart from PDCP Service Data Unit (SDU)s, the $Xw$ interface is also used for flow control feedback. The $Xw$-$U$ interface is used to deliver LWA PDUs from eNB to WT. The eNB-WT control plane signalling for LWA is performed by means of $Xw$-$C$ interface signalling. It supports the following functions: \emph{i)} Transfer of WLAN metrics from WT to eNB, \emph{ii)} Support of LWA for UE (establishment, management and control of user plane), and \emph{iii)} $Xw$ management and error handling function. Figure \chg{\ref{fig1:sub2}} shows the protocol architecture of user plane for the non-collocated LWA approach. It is defined to allow connection of the LTE network to the already existing Wi-Fi deployments (i.e. Wi-Fi APs can operate as native APs while not forwarding the LWA traffic).

A new sublayer, called LWA Adaptation Protocol (LWAAP), that adds the Data Radio Bearer (DRB) ID to the PDCP frames and transmits it to the Wi-Fi interface is defined. This allows multiple bearers to be offloaded to the Wi-Fi network. In the control plane, the eNB is responsible for selecting the bearers to offload to WLAN and for the activation/deactivation of LWA. However, Release 13 does not specify any algorithm for the interface selection.
Regardless of the deployment scenario, PDCP frames are scheduled by the eNB, where some frames are encapsulated with Wi-Fi protocol and transmitted through the Wi-Fi interface. \chg{LWA can also configure the network to allow use of both Wi-Fi and LTE simultaneously. This procedure, called split bearer, uses both eNB and Wi-Fi radio resources. On the contrary, LWA also allows switched bearers, which only utilizes the WLAN radio resources for transmission of frames from the eNB to the UE. 

While using the split bearer operation, the UE supports in-sequence delivery of frames based on the reordering procedure introduced in 3GPP Release 12 for dual connectivity. This is done using a reordering window controlled by a reordering timer running at the UE. The received frames from both LTE and WiFi are reordered by the aforementioned additional PDCP functionality of the LWA UE.
In the uplink, PDCP PDUs can only be sent via the LTE. }
\section{Overview of LWIP architecture}
\label{Overview of LWIP architecture}
\begin{figure}
\centering
\begin{subfigure}[b]{.5\textwidth}
  {\raisebox{10mm}{
  \includegraphics[width=1\linewidth]{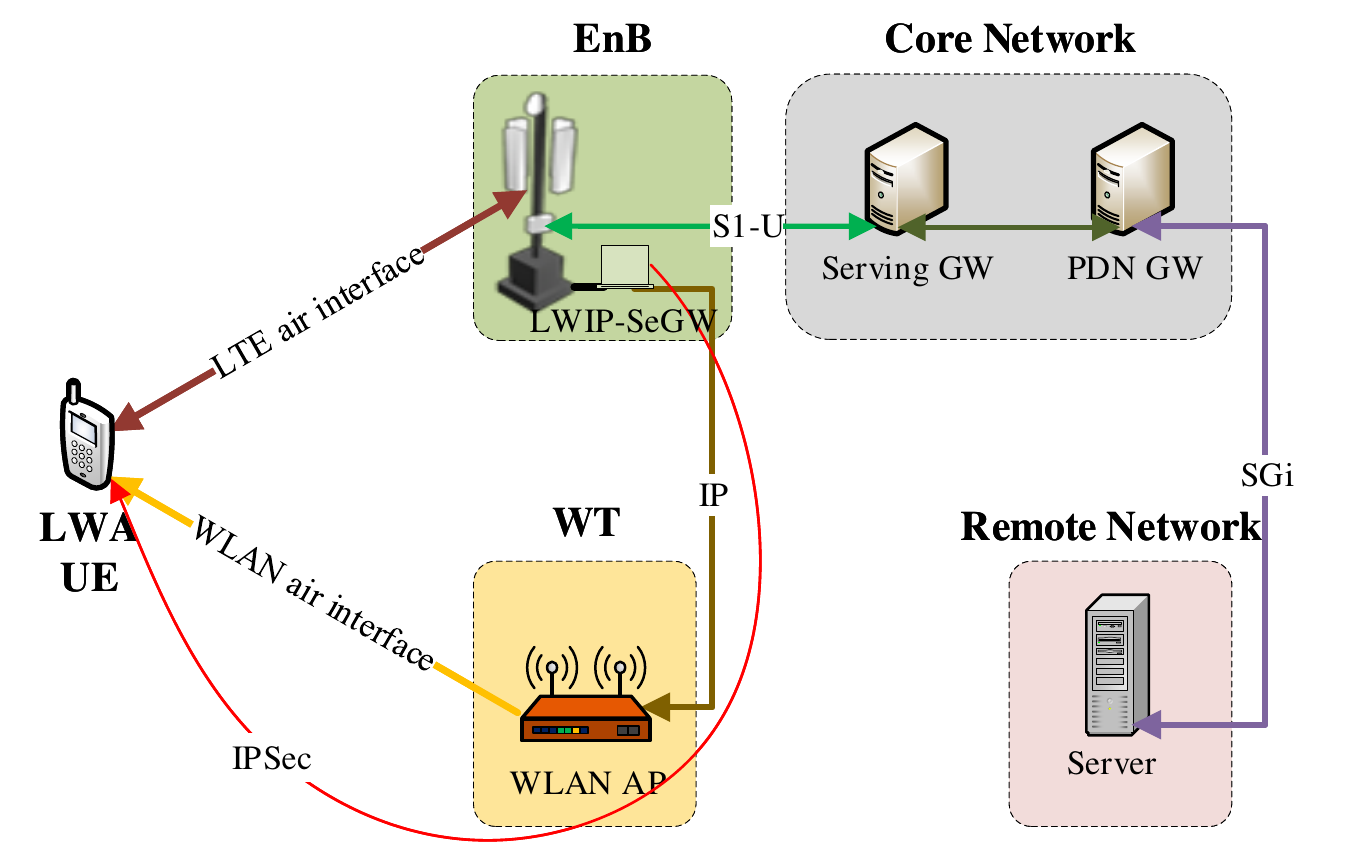}}}
  \caption{Network Architecture.}
  \label{fig2:sub1}
\end{subfigure}%
\begin{subfigure}[b]{.5\textwidth}
  \centering
  \includegraphics[width=1\linewidth]{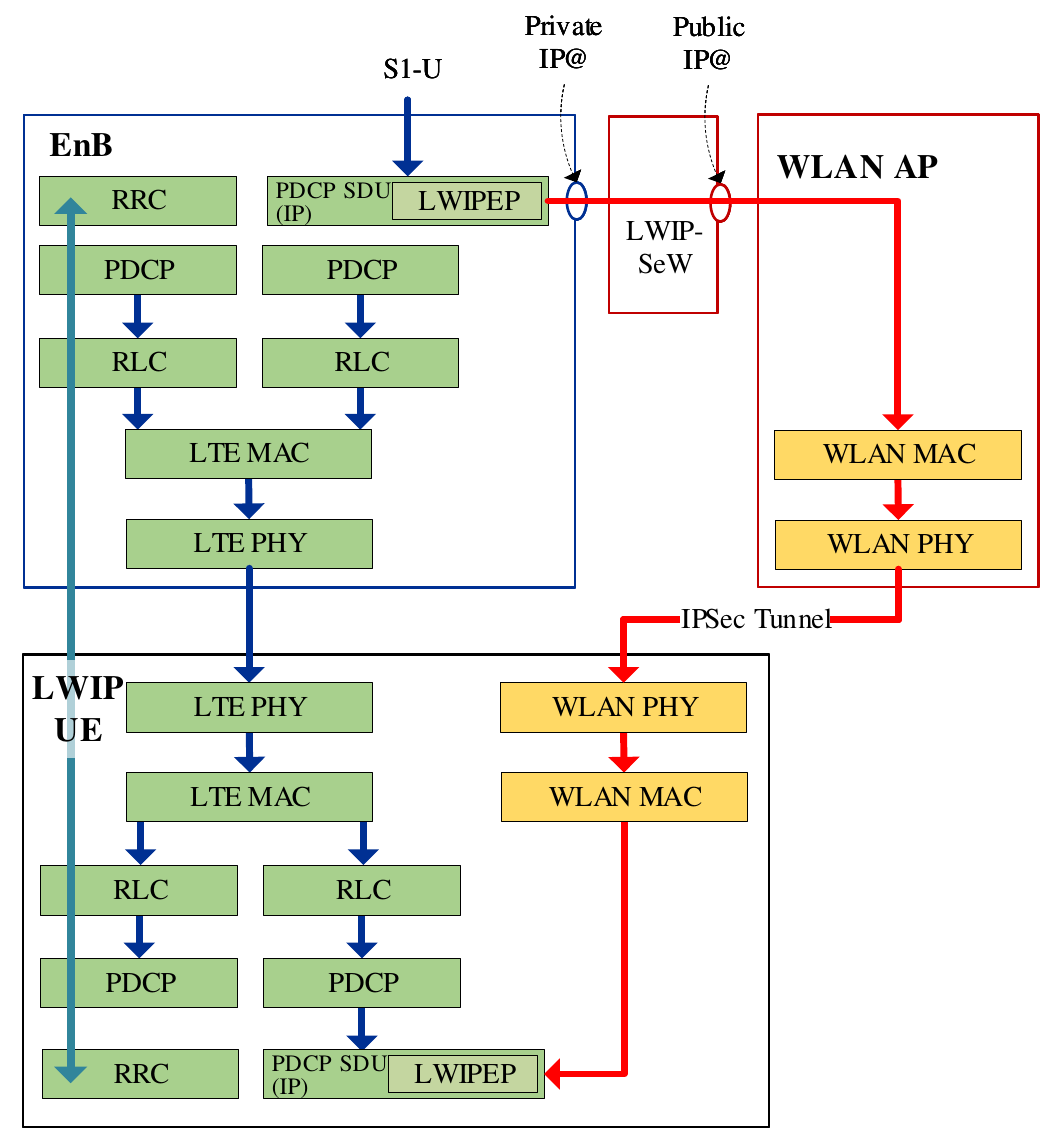}
  \caption{Protocol architecture defined by 3GPP.}
  \label{fig2:sub2}
\end{subfigure}
\caption{Overview of LWIP.}
\label{fig2}
\end{figure}
The foundation of 3GPP LWIP release 13 was set by Wi-Fi boost, which realized the first implementation of downlink switching of Internet Protocol (IP) layer connectivity. The basic aim of LWIP was to readily adopt and supplement LTE without making any major changes to the WLAN infrastructure. Figure \ref{fig2:sub1} shows the network architecture of LWIP scheme. 

Figure \ref{fig2:sub2} shows the protocol architecture for LWIP. As can be seen, traffic splitting is performed above the PDCP layer, and data path bypasses all LTE protocols. The LWIP aggregation scheme involves the use of an Internet Protocol Security (IPSec) tunnel to transfer PDCP SDU packets from eNB to UE via WLAN infrastructure. IPSec is an end-to-end security framework (which includes a set of protocols and algorithms for mutual authentication) that operates at the network layer by extending the IP header of the frames. That is, in the IPSec tunnel mode, the entire IP packet is protected by inner and outer IP datagrams. The inner headers specify the communication endpoints whereas the outer IP header defines cryptographic end-points. Since the IPSec connection is unconcerned of WLAN deployments, the LWIP solution supports legacy WLAN deployments more easily than LWA (that requires software as well has hardware additions).

While the LWIP scheme is used, the Radio Resource Control (RRC) and signalling messages, which are exchanged between eNB and the UE, are carried over the LTE interface. A new protocol, called LWIP Encapsulation Protocol (LWIPEP), is specified for the transfer of user plane data. LWIPEP is also used for identification of the data bearer identity to which the LWIPEP SDU belongs. The traffic splitting is performed above the PDCP layer and the data path going towards WLAN is only passed through LWIPEP layer (and is bypassed by the LTE protocols below the PDCP layer). The LWIPEP protocol allows forwarding frames of different DRB using the IPSec tunnel configured for the UE. It is important to note here that no reordering procedure is employed at the UE (since aggregation is done above the PDCP layer) and all LWIP bearer data is forwarded towards to Wi/Fi network. 

Since LWIP is transparent to the WLAN, the flow control mechanism used in LWA are not applicable. Also, the $Xw$ interface and WT nodes are not required. However, due to security issues, the IPSec tunnel is terminated into a dedicated gateway, called Security Gateway (SeGW). It is an interconnection point between WLAN and eNB (with a dedicated interface) and can be deployed at the eNB. Due to LWIP SeGW, the UE does not have a direct IP connectivity to the eNB through WLAN. Each UE is assigned a unique IPSec tunnel.

The eNB is responsible for activating and deactivating LWIP operations based on UE measurement reporting. The main drawback of this scheme is that the IPSec tunneling protocol appends an IPSec header to the Downlink IP packets that travel from eNB to UE via the WLAN network.

\section{Comparison between LWA and LWIP}
\chg{In both LWA and LWIP, the licensed operation (with aggregated LTE link) provides coverage and mobility robustness, whereas, the unlicensed operation (with aggregated WLAN link) provides higher (peak) UE data throughput by routing of high data traffic. There are a number of similarities between LWA and LWIP. In more detail: \emph{i)} both techniques are controlled/activated by the eNB, \emph{ii)} similar Wi-Fi measurement reporting frameworks are used in both,  \emph{iii)} the same mobility set is used to determine the AP to which the UE would be connected to and \emph{iv)} regular Wi-Fi service can run in parallel to these techniques over the same hardware. In terms of security architecture, both the techniques use WLAN authentication methods to provide secure authentication and protection against relay attacks. 

\begin{table}[H]
	\caption{Comparison of LWA and LWIP across different performance indicators.}
	\label{tab:1}
	\centering
	{
    	\scalebox{0.9}{
		\begin{tabular}{|>{\centering\arraybackslash}p{6cm}|>{\centering\arraybackslash}p{3cm}|>{\centering\arraybackslash}p{3cm}|}
			\hline
			\parbox[c][5ex]{5ex}{\centering}\bf{Attributes}&\bf{LWA}&\bf{LWIP}\\ \hline
			\parbox[c][4ex]{4ex}{\centering}Specifications & 3GPP Release 13&3GPP Release 13 \\\hline
			\parbox[c][4ex]{4ex}{\centering}eNB control&Yes&Yes  \\\hline       
			\parbox[c][4ex]{4ex}{\centering}Connecting layers&PDCP&IP (PDCP SDU) \\\hline
			\parbox[c][4ex]{4ex}{\centering}Offloading granularity&Split or Switched Bearer & Switched Bearer\\\hline
			\parbox[c][4ex]{4ex}{\centering}Modification to legacy Wi-Fi AP&Yes&No  \\\hline  
			\parbox[c][4ex]{4ex}{\centering}Development scenarios&Collocated or non-collocated&Collocated\\\hline  
			\parbox[c][4ex]{4ex}{\centering}Upgrade in LTE network&eNB and UE&eNB\\\hline  
            \parbox[c][4ex]{4ex}{\centering}Upgrade in Wi-Fi network&LWA cognizant AP&None\\\hline
            \parbox[c][4ex]{4ex}{\centering}New network entities in LTE&LWAAP and $Xw$ Interface&LWIPEP and SeGW\\\hline
            \parbox[c][4ex]{4ex}{\centering}New network nodes in Wi-Fi&WT&None\\\hline
            \parbox[c][4ex]{4ex}{\centering}Additional interface for flow control and security&$Xw$&None\\\hline
            \parbox[c][4ex]{4ex}{\centering}WLAN measurements&Yes&Yes\\\hline
            \parbox[c][4ex]{4ex}{\centering}WLAN security&WLAN native 802.1x EAP/AKA with fast authentication method&WLAN native 802.1x EAP/AKA\\\hline
            \parbox[c][4ex]{4ex}{\centering}Regular Wi-Fi services on the same hardware&Yes&Yes\\\hline
            \parbox[c][4ex]{4ex}{\centering}Additional UE cost&New software&New software\\\hline
            \parbox[c][4ex]{4ex}{\centering}WLAN traffic direction&Downlink&Downlink plus uplink\\\hline
		\end{tabular}
	}}
\end{table}

Despite the similarities, LWA and LWIP techniques have significant differences. The main difference between these two architectures is the protocol layers at which the aggregation occurs and the security mechanism applied. In LWA, the eNB forwards PDCP data units (PDUs) to the WLAN AP and in LWIP, the eNB forwards PDCP SDUs, i.e., IP packets, to the WLAN AP. Table \ref{tab:1} provides a comparison between LWA and LWIP in terms of some of the key performance indicators.    

The advantage of LWA is that it can provide better control and utilization of radio resources on both LTE and Wi-Fi links (but requires careful splitting of the traffic between LTE and Wi-Fi to ensure good load balance). This, in return,  can increase the aggregate throughput for all users and improve the total system capacity. The non-collocated case of LWA allows higher density, independent placement, and, most importantly leverages the existing WLAN deployments. 

While LWA in collocated scenario can be enabled via a simple software upgrade to LTE and Wi-Fi deployments and doesn't require deploying all new hardware with the associated cost, LWIP is a solution that is agnostic to the WLAN infrastructure (i.e. the IPSec tunnel is transparent to WLAN). LWA non-collocated flow control mechanisms are not applicable in LWIP. For this reason, the $Xw$ interface and the WT node are not required (but can be used to implement LWIP). In LWIP, eNB is in control of traffic steering. This protocol shows compatibility with any legacy WLAN, as it does not need any modifications or enhancements to the Wi-Fi AP. The drawbacks of LWIP are that the eNB hosts the SeGW (which is a new component) and it can only support switched bearer option (where the capacity of both LTE and Wi-Fi networks can not be used together).}


\section{Building Blocks}
\label{Background}
As highlighted in the previous sections, the basic aim of this work is to extend NS-3 to incorporate the additional LWA and LWIP functionalities to enable the support of LTE transmissions in the unlicensed band. In this section, we first explore the basics of the NS-3 simulator. It is important to mention here that NS-3 has Wi-Fi and LTE modules already included, but these technologies were adapted (in particular, LTE PDCP layer was updated) and used together to enable the LWA and LWIP implementation.

NS-3 is an open-source, discrete event-based network simulator, which is licensed under the GNU GPLv2 license. It is currently one of the most powerful and complete network simulators with a strong community support. It aims to provide open and extensible representative models of real world communication networks so as to produce realistic behaviors and statistics. 

\begin{figure}
\centering
\begin{subfigure}{.5\textwidth}
  \centering
  \includegraphics[width=1\linewidth]{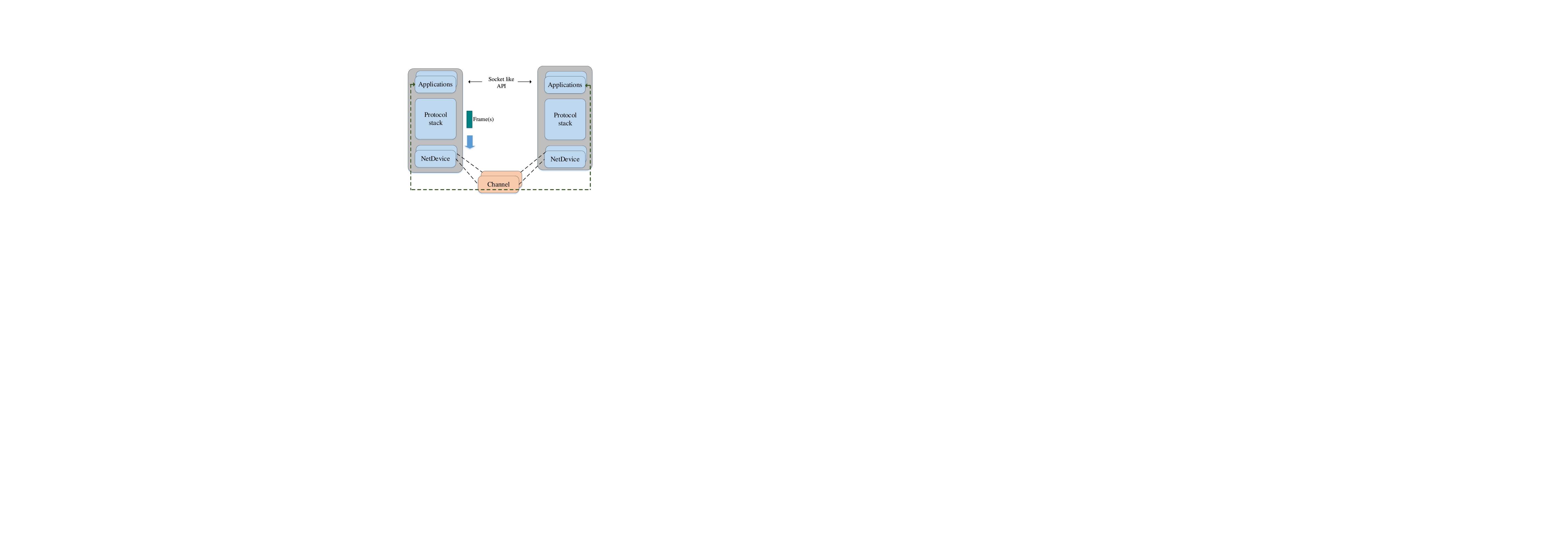}
  \caption{Fundamental Objects}
  \label{fig:sub1}
\end{subfigure}%
\begin{subfigure}{.5\textwidth}
  \centering
  \includegraphics[width=1\linewidth]{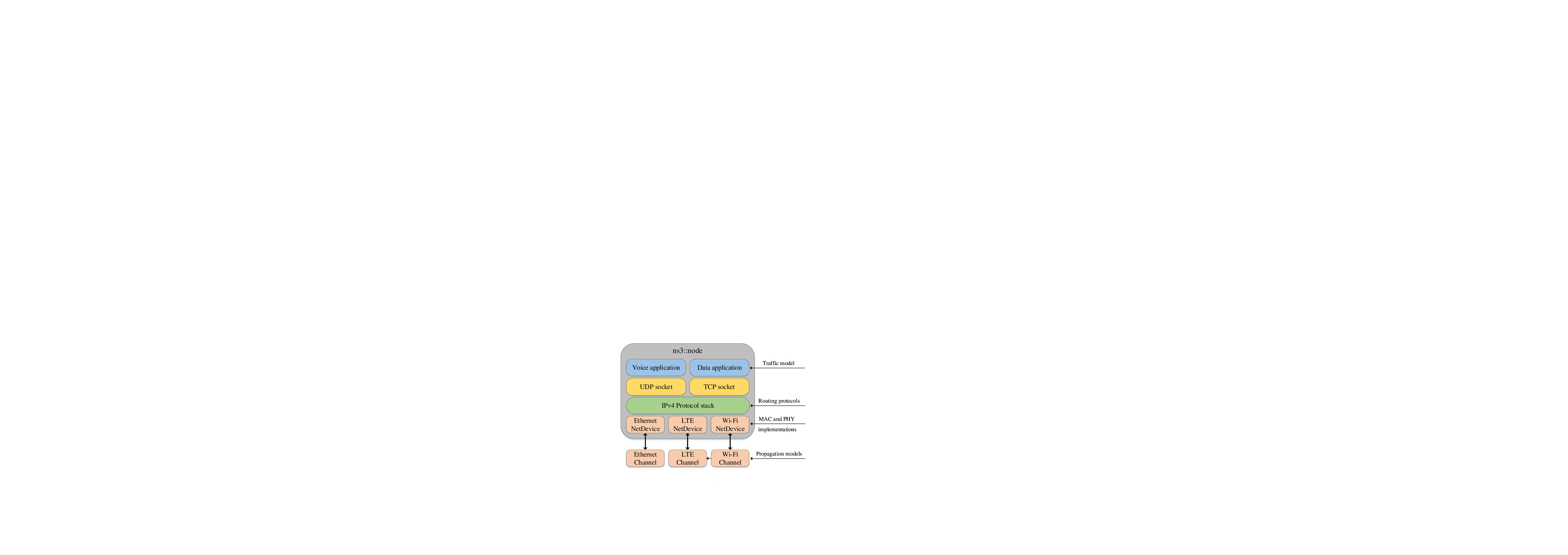}
  \caption{Overview of a node model}
  \label{fig:sub2}
\end{subfigure}
\caption{NS-3 overview}
\label{fig:test}
\end{figure}
The key networking abstraction used by NS-3 comprises node, application, channel, net device (NetDevice) and topology helpers. As shown in Figure \ref{fig:sub2}, a node is the basic computing device which contains different applications, protocol stacks and software drivers plus simulated hardware (representing a network interface card). Application framework provides methods for managing representation of user-level applications within simulations. A node is connected to a channel which is a logical path over which information flows. It provides methods to manage communication sub-networks and connects nodes over them. Net device represents the network device card which connects a node to a network. In NS-3, the net device abstraction covers both the software driver and the simulated hardware. A net device is installed in a node in order to enable it to communicate with other nodes within the simulation through different channels. Nodes can also contain multiple net devices and application objects. Examples of different applications available in NS-3 are as follows: OnOff, PacketSink, UdpEcho, BulkSend, UdpClientServer etc. Sockets (UDP/TCP) are then used to port the existing application to the NS-3 environment. To allow an application to operate over two nodes, one node is set as a source while the other node is set as the receiver. These sockets can be identified using IPv4/IPv6 addresses. Receiver socket is used to bind with the incoming request from the socket and source socket connects to the receiver socket by identifying it with a given IP address. For synchronization purposes, the port number of both the source and receiver sockets is kept the same.  

NS-3 allows a programmer to follow specific simulation events by creating new tracing functions. In addition, different network statistics can be inspected using already existing network monitoring statistics, called FlowMonitor. The flow metrics, such as throughput, delay, jitter and packet loss ratio can be observed using FlowMonitor. NS-3 also allows the possibility to store multiple layer output events in a text file and to trace packet transmit/receive events via Packet Capture (PCAP) files. 

NS-3 consists of a set of libraries and other external software libraries that can be combined together to evaluate a variety of access technologies. It supports the development of new components using C++ or Python. In our work we opted to use C++. To the best of our knowledge, none of the existing work has investigated the development and evaluation of LWA and LWIP protocols in NS-3.

\subsection{LTE implementation in NS-3}
\label{LTE implementation in NS-3}
\begin{figure}[tb]
	\centering
	\includegraphics[width=0.7\textwidth]{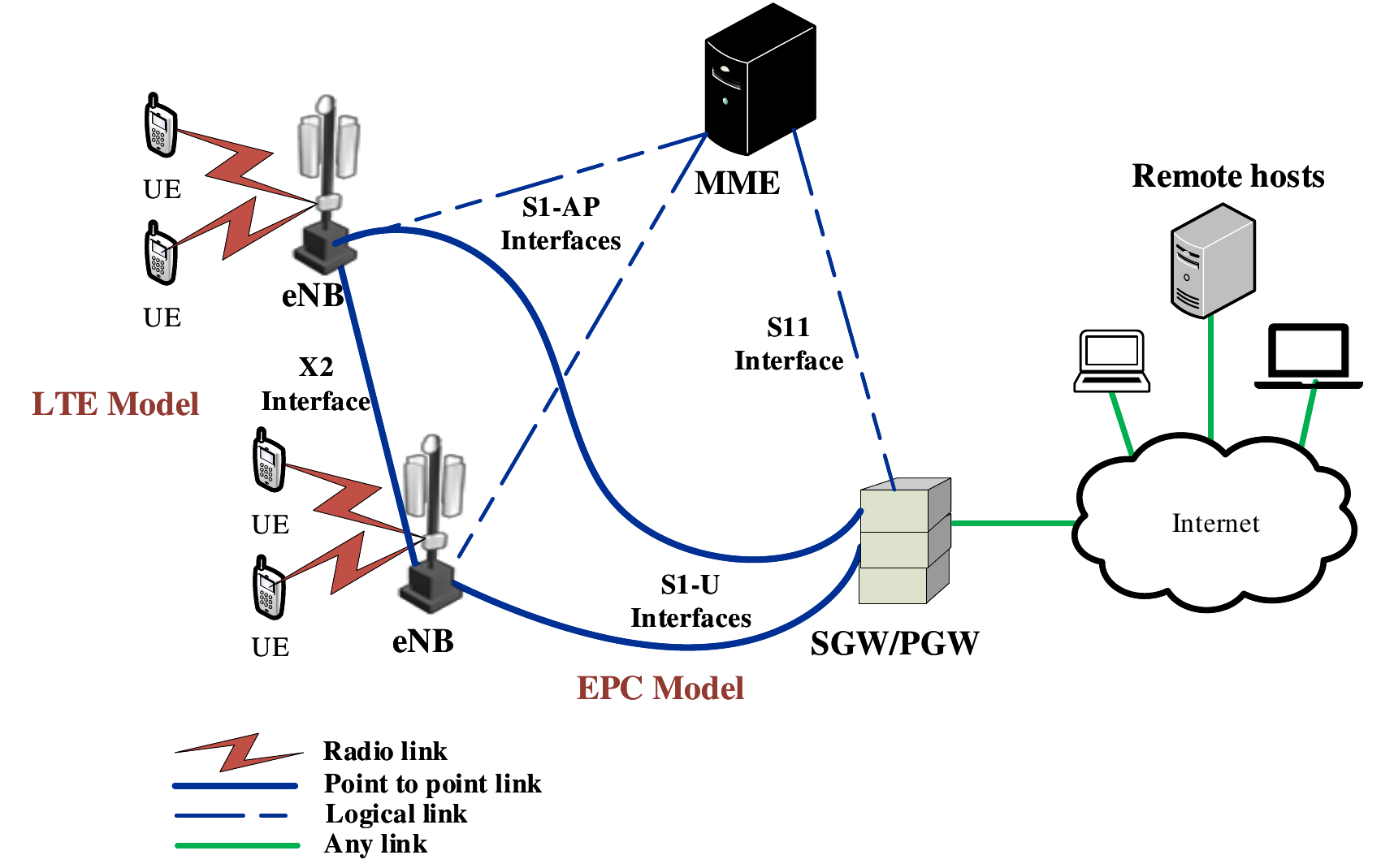}
	\caption{LTE EPC model.}
	\label{fig2a}
\end{figure}
We focus our work in the use of the NS-3 LTE module called LENA (LTE-EPC Network Simulator)\cite{LTEEPCLENA}, which is developed and maintained by the Centre Tecnologic de Telecomunicacions de Catalunya (CTTC), Barcelona, Spain. As shown by Figure \ref{fig2a}, the LTE-EPC model has two main components; LTE module and Evolved Packet Core (EPC) model. The LTE model represented by pink, includes protocol stacks such as Radio Resource Control (RRC), PDCP, Radio Link Control (RLC), Medium Access Control (MAC) and Physical (PHY) layer. It also handles and manages the communication between UEs and eNB station. The EPC Model, represented by blue lines and green lines, contains the communication between different network interfaces.

The LTE module in NS-3 has been designed to support the evaluation of several aspects of LTE elements, namely Radio Resource Management (RRM), QoS-aware packet scheduling, intercell interference coordination and dynamic spectrum access. It includes the LTE Radio Protocol stack (PDCP, RLC, MAC, PHY) which resides entirely within the eNB and the UE.

\chg{The EPC model used in LTE NS-3 is depicted in Figure \ref{fig2b}. It provides the means to simulate end-to-end IP connectivity over the LTE Model. The remote host is connected to the LTE network through the Service Gateway/Packet Gateway (SGW/PGW) node of this model. Two different layers of IP networking are defined in the EPC model. The first one is the IP networking of EPC local area network. It involves the connection of eNB and SGW/PGW nodes through a set of point-to-point links. The second IP networking is the end-to-end connectivity for end users that consists of UEs, the SGW/PGW and the remote host. It does not include the eNB. 

Based on Figure \ref{fig2b}, the downlink data flow is explained as follows: IP packets are generated at the remote host and addressed to the UE device. The internet routing algorithm diverts these packets to the netdevice at the PGW/SGW node. A virtual net device is installed at the PGW/SGW node, which is assigned the gateway IP address of the UE subnet. Thus, the static routing rules will allow the incoming frames to be routed through this Virtual net device. Afterwards, packets are sent to the eNB through the S1-U interface. Upon receiving the packets, the RLC and PDCP instances of eNB net device are used to retrieve the Radio Bearer ID (RBID) information embedded in a tag attached with the packets. This RBID information is used to forward the packet over the LTE radio interface. Finally, the UE net device receives the frame and forwards it locally to the IP stack.
\begin{figure}[htb]
	\centering
	\includegraphics[width=0.55\textwidth]{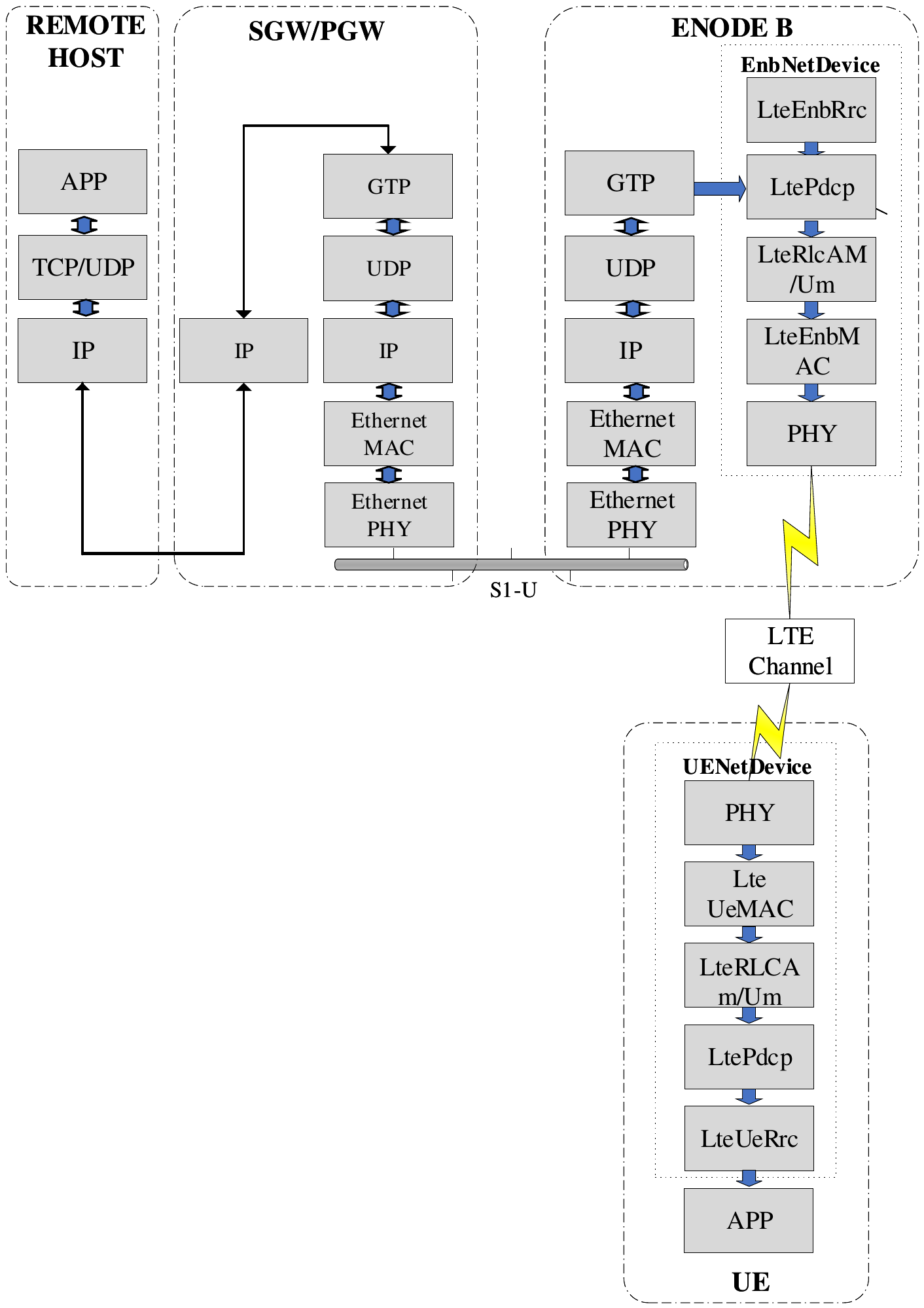}
	\caption{LTE module architecture in NS-3.}
	\label{fig2b}
\end{figure}
The LTE network allows data traffic from the eNB to be routed to the remote host or UE through the EPC and handles the management and signaling functions of LTE. An important aspect of this implementation is that a secure tunnel is created from the PGW/SGW node, through eNB to the UE. In addition, the static routing algorithm diverts traffic from/to the UE. Therefore, the second IP networking does not involve the use of IP stack at the eNB in the EPC model of the NS-3.}

In this work, we focus on extracting data from either the PDCP layer, of the eNB module in NS-3, so that it could be offloaded to a Wi-Fi infrastructure network.
\subsubsection{NS-3 PDCP layer implementation}
\chg{The main function of PDCP layer in LTE-EPC model is to allow transmission of both data and signalling in a unified manner. Apart from this, PDCP layer also manages the compression and decompression of both the header and the content of IP data packets. It can also manage integrity protection and verification in the control plane. It is responsible for handling bearers and Each PCDP PDU can be uniquely by the following fields: 
\begin{itemize}
    \item PCDP SN
     Each PDCP PDU is identified by a Sequence Number (SN), which is put in the PDCP header. It is used for in-sequence delivery of upper layer PDUs and to eliminate duplicated lower layer SDU.
    \item LCID
    A Logical Channel Identifier (LCID) is also used to identify the different logical channels corresponding to different QoS requirements.
    \item RNTI
     A Radio Network Temporary Identifier (RNTI) is assigned by the eNB to identify the UE over the radio interface.
\end{itemize}
With respect to 3GPP specification, The PDCP model of NS-3 supports the following features:
\begin{enumerate}
    \item Transfer of data (user plane or control plane),
    $DoTransmitPdcpSdu()$ interface is provided to upper RRC entity to transmit. The PDCP entity calls this primitive to send data to RLC SAP Provider (i.e. the part of the SAP that contains the RLC methods called by the PDCP). $DoReceivePdu()$ is the interface provided to lower RLC entity.  The PDCP entity calls this primitive to receive  entity to notify the RRC entity of the reception of a new RRC PDU.  
    \item Maintenance of PDCP SNs,
    Status variables of the PDCP are maintained by TX sequence number $txSn$ and RX sequence number $rxSn$. 
    \item Transfer of SN status (for use upon handover).
    $GetStatus()$ is used to get the status of the PDCP that includes the sequence number of the received and transmitted frames.
\end{enumerate}
Other PDCP features, such as header compression and ciphering are not implemented. The trace functionality available at the PDCP of the current NS-3 LTE implementation includes LCID, RNTI and packet size information for frames to be transmitted to  transmitted and received. 

In order to implement the LWA and LWIP schemes, the PDCP layer was enhanced to allow the possibility to forwarded data to the primary RLC layer or to offload it to the Wi-Fi network. This was achieved by passing a state variable $lwaactivate$ to the PDCP layer. The detail operation of the aforementioned variable are provided in Section~\ref{Packet generation at remote host and transfer to PDCP layer}.}

\subsubsection{NS-3 LTE code}
In this section, we describe a simple LTE/EPC example program. This script instantiates one eNB and attaches a single UE to it. It also creates an EPC network which is connected to a remote host. The three main component of EPC core in LTE (i.e. SGW, PGW and MME) are implemented in the following NS-3 code snippet.
\begin{lstlisting}[numbers=left, breaklines=true]
Ptr<LteHelper> lteHelper = CreateObject<LteHelper> ();
Ptr<Node> pgw = epcHelper->GetPgwNode ();
Ptr<PointToPointEpcHelper>  epcHelper = CreateObject<PointToPointEpcHelper> ();
lteHelper->SetEpcHelper (epcHelper);
\end{lstlisting}
The statements in line 1 and 2 create an LTE object and SGW/PGW node. The last two statements interconnect them together. The interconnection between them is made automatically by the LteHelper object. The eNB and UE are created and attached together by using the following code.
\begin{lstlisting}[numbers=left, breaklines=true]
NodeContainer ueNodes;
NodeContainer eNBNodes;
eNBNodes.Create(1);
ueNodes.Create(1);
NetDeviceContainer eNBLteDevs = lteHelper->InstalleNBDevice (eNBNodes);
NetDeviceContainer ueLteDevs = lteHelper->InstallUeDevice (ueNodes);
lteHelper->Attach (ueLteDevs, eNBLteDevs);
\end{lstlisting}
Line 1 to 4 are used in creating a single eNB node and a UE. The 4 and 5 line create net devices at the eNB and UE. The line 7 attaches the UE with the eNB via their respective net devices.
The following code outlines how to connect a single remote host to the PGW via a point-to-point link.
\begin{lstlisting}[numbers=left, breaklines=true]
NodeContainer remoteHostContainer;
remoteHostContainer.Create (1);
Ptr<Node> remoteHost = remoteHostContainer.Get (0);
InternetStackHelper internetLTE;
internetLTE.Install (remoteHostContainer);

PointToPointHelper p2ph;
p2ph.SetDeviceAttribute ("DataRate", DataRateValue (DataRate ("100Gb/s")));
p2ph.SetDeviceAttribute ("Mtu", UintegerValue (1500));
p2ph.SetChannelAttribute ("Delay", TimeValue (Seconds (0.010)));
NetDeviceContainer internetDevices = p2ph.Install (pgw, remoteHost);
Ipv4AddressHelper ipv4h;
ipv4h.SetBase ("10.0.0.0", "255.0.0.0");
Ipv4InterfaceContainer internetIpIfaces = ipv4h.Assign (internetDevices);
\end{lstlisting}
	The following code enables the remote host to reach the UE by specifying static routes. The PointToPointEpcHelper, by default, assigns an IP address of the 7.0.0.0 network to the UE .
\begin{lstlisting}[numbers=left, breaklines=true]
Ipv4StaticRoutingHelper ipv4RoutingHelper;
Ptr<Ipv4StaticRouting> remoteHostStaticRouting =
ipv4RoutingHelper.GetStaticRouting (remoteHost->GetObject<Ipv4> ());
remoteHostStaticRouting->AddNetworkRouteTo (Ipv4Address ("7.0.0.0"), Ipv4Mask ("255.0.0.0"), 1);
 \end{lstlisting}
\subsection{Wi-Fi implementation in NS-3}
\label{Wi-Fi implementation in NS-3}
\begin{figure}[htb]
	\centering
	\includegraphics[width=1\textwidth]{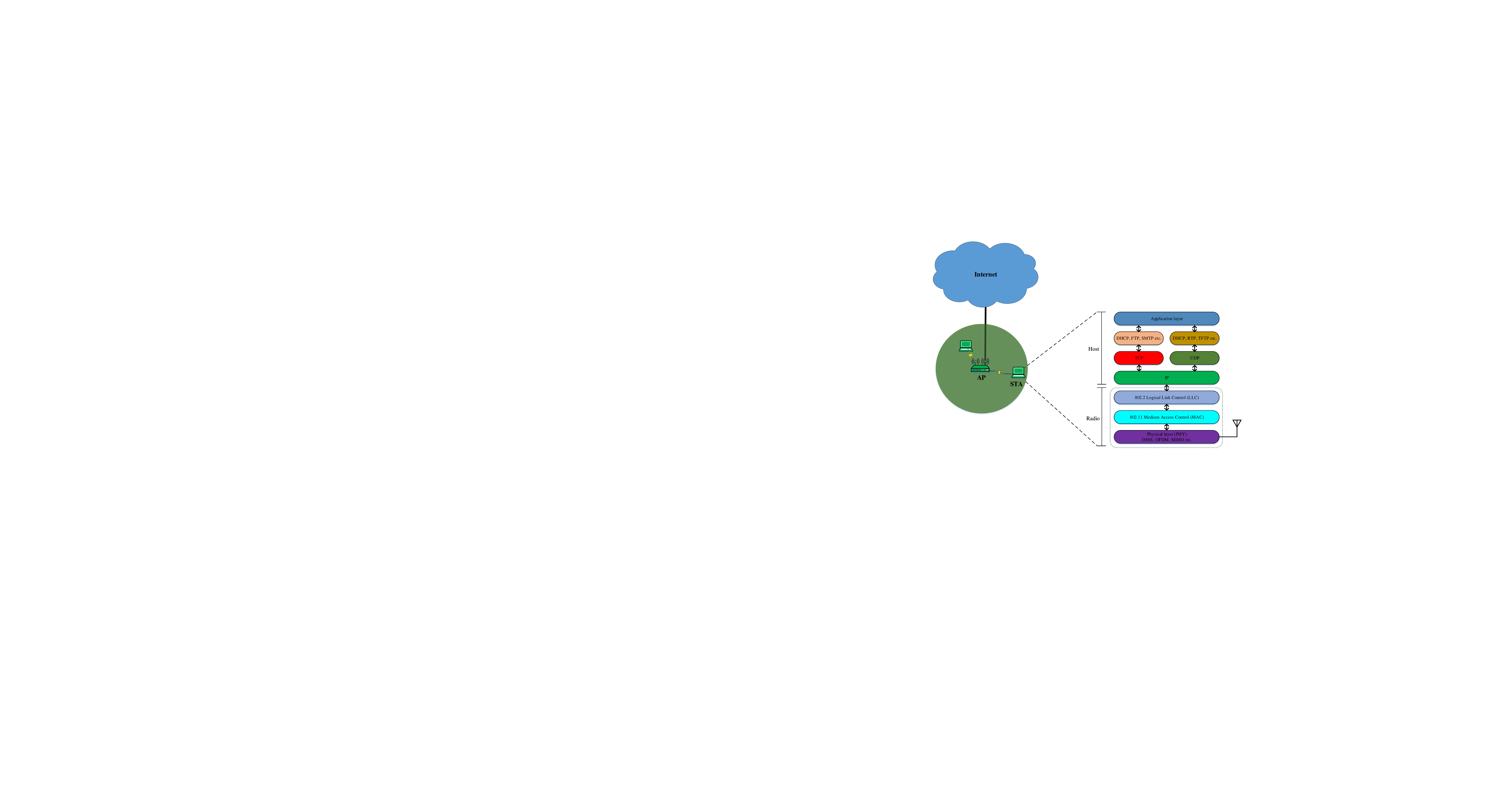}
	\caption{IEEE 802.11 protocol layers.}
	\label{fig3aa}
\end{figure}

Wi-Fi technology is based on the IEEE 802.11 standard that defines the MAC and PHY layer for implementation of WLAN communication. Figure \ref{fig3aa} highlights the fundamental architecture of Wi-Fi in infrastructure mode. The basic building block of a Wi-Fi network is called the Basic Service Set (BSS). It is the area or cell that an AP covers (i.e. the transmissions from the AP have signal to noise ratio above the decodable threshold within this range) and all non-AP stations\footnote{In this report, we use the term "station" to represent a non-AP station.} communicate in a centralized manner to the AP.

The MAC layer of 802.11 consists of the Distributed Coordination Function (DCF) and the Point Coordination Function (PCF), which are used to coordinate stations that perform simultaneous data transmissions. PCF, a central polling scheme in 802.11, is not mandatory in the IEEE 802.11 specifications. DCF is the dominant protocol used due to its simple and distributed implementation. It operates by using the well-known carrier sense paradigm, with an exponential backoff mechanism devised to ensure low probability of simultaneous transmission attempts by multiple stations. It also guarantees the same probability of channel access for all the stations intending to transmit over the shared channel.

The PHY layer specifications of IEEE 802.11 concentrate mainly on wireless transmission and on performing secondary functions, such as assessing the state of the wireless medium and reporting it back to the MAC layer. It is divided into two sublayers: Physical Layer Convergence Procedure (PLCP) and Physical Medium (PMD). The MAC layer communicates to the PLCP sublayer through Service Access Points (SAP). The IEEE 802.11 Physical carrier sensing includes the Physical Clear Channel Access Mechanism (PHYCCA) procedure, which  is a function of the PLCP sublayer. The actual transmission and reception of data is performed at the PMD layer under the directions of the PLCP sublayer.

The Wi-Fi module in NS-3 is by far the largest network device component. It contains numerous modules and sub-classes and implements the IEEE 802.11 MAC and Physical layers that conform to the 802.11a, 802.11b, 802.11g, 802.11n and 802.11ac specifications.
\begin{figure}[tb]
	\centering
	\includegraphics[width=0.5\textwidth]{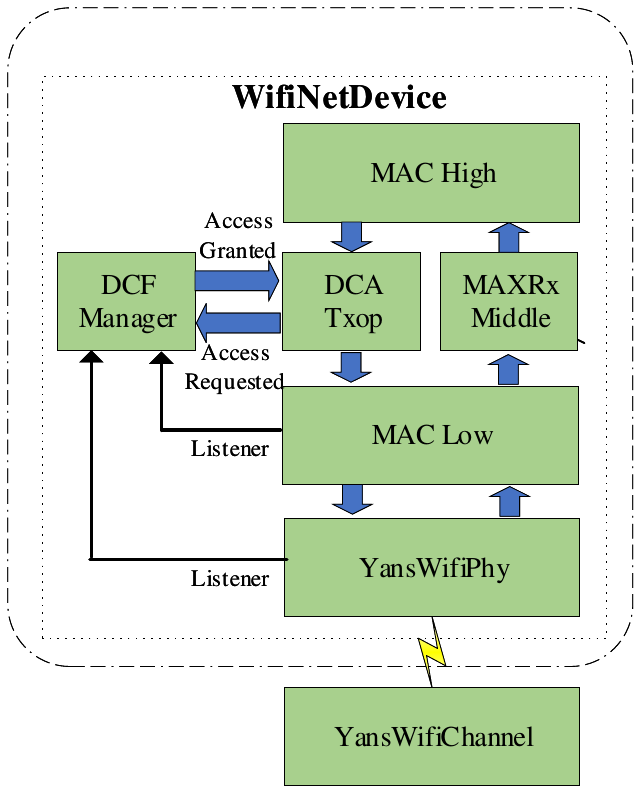}
	\caption{Wi-Fi models and interactions in NS-3.}
	\label{fig4a}
\end{figure}
A Wi-Fi net device is built up of classes that coordinate the packet reception and transmission. As highlighted in Figure \ref{fig4a}, the main components of NS-3 Wi-Fi implementation consist of the following modules: WifiNetDevice, MacHigh, DcfManager, DcaTxop, MacRx, MacLow, YansWi-FiPhy, and YansWi-Fichannel. 

The WifiNetDevice module is used for interfacing with higher layers. The MacHigh models management functionalities such as beacon generation, probing and association frame exchange. Two types of MAC high classes are available for infrastructure configuration in NS-3. Namely, $ns3 :: ApWifiMac$ (AP class) and $ns3 :: StaWifiMac$ (station class). The AP class models functions of an AP such as beacon generation, association and authentication. In addition, it performs all the necessary methods to receive, process and forward packets to the lower and higher layers. The different MACHigh classes share a common parent class $ns3:: RegularWifiMac$ that implements the basic features related to QoS and provides support for High Throughput (HT) and Very High Throughput (VHT) wireless configuration.
MacHigh forwards incoming data to a single DcaTxop. The main classes implementing the DCF mechanism for the MAC layer are DcfManager and DcaTxop classes. DcaTxop contains the backoff counter and pushes data to MacLow for transmission after gaining medium access from the DcfManager. DxaTxop and MacRXMiddle together handle the sequence number of packets, fragmentation, filtering of duplicate reception, and re-transmissions. MacLow module takes care of the Request to Send (RTS)/Clear to Send (CTS)/Data and Acknowledge (ACK) packet transmissions.

The $ns3::YansWifiPhy$ class is designed to implement the IEEE 802.11 physical layer. It models an additive Gaussian Noise Channel (AWGN) with cumulative noise handled by InterferenceHelper and implements a packet level physical model. It manages the reception and transmission of the packets over the wireless channel, models channel interference and errors, channel sensing, Signal to Interference plus Noise Ratio (SINR) computation, etc. In addition, this class allows the tracking of the energy consumed during the transmissions and also while listening to the channel. The abstract class $ns3:: YansWifiChannel$ provides an analytical approximation of the physical medium over which the Wi-Fi data is transmitted. It permits the definition of the propagation loss and propagation delay models of the channel. The delay of the received frames is calculated according to the PropagationDelayModel class and the calculation of the reception power is performed by the PropagationLossModel class.  

\subsubsection{NS-3 Wi-Fi Infrastructure mode code} 
In the following section, a simple description is provided to design a simulation model of Wi-Fi network in an infrastructure mode within the NS-3 environment. We start by first declaring two Wi-Fi stations (i.e. a single AP and a single station)
\begin{lstlisting}[numbers=left, breaklines=true]
NodeContainer WifiStaNodes;
NodeContainer WifiApNode;
WifiApNode.Create(1);
WifiStaNodes.Create(1);
\end{lstlisting}
The Wi-Fi standard properties can be defined by WifiHelperclass with the following code:
\begin{lstlisting}[numbers=left, breaklines=true]
WifiHelper Wifi;
Wifi.SetStandard (Wifi_PHY_STANDARD_80211ac);
StringValue DataRateWifi = VhtWifiMacHelper::DataRateForMcs (datamcs);
StringValue ControlRateWifi
VhtWifiMacHelper::DataRateForMcs (controlmcs);
Wi-Fi.SetRemoteStationManager ("ns3::ConstantRateWi-FiManager", "DataMode", DataRateWifi,"ControlMode", ControlRateWifi);
\end{lstlisting} 
 Next, the details of the Wi-Fi channel to be used in the simulations are added.
 \begin{lstlisting}[numbers=left, breaklines=true]
YansWifiChannelHelper WifiChannel;
WifiChannel.SetPropagationDelay ("ns3::ConstantSpeedPropagationDelayModel");
WifiChannel.AddPropagationLoss
("ns3::FixedRssLossModel","Rss",DoubleValue (rss)); WifiPhy.SetChannel (WifiChannel.Create ());
\end{lstlisting}                          
Next, MAC and PHY protocols are installed on the Wi-Fi station and the AP.
 \begin{lstlisting}[numbers=left, breaklines=true]
YansWifiPhyHelper WifiPhy =  YansWifiPhyHelper::Default ();
VhtWifiMacHelper WifiMac = VhtWifiMacHelper::Default ();
NetDeviceContainer devicesta;
Ssid ssid = Ssid ("ns3-80211ac");
WifiMac.SetType ("ns3::StaWi-FiMac", "Ssid", SsidValue (ssid), "ActiveProbing", BooleanValue (false));
devicesta.Add(Wi-Fi.Install (WifiPhy, WifiMac, WifiStaNodes)); 
WifiMac.SetType ("ns3::ApWifiMac","Ssid", SsidValue (ssid)); 
NetDeviceContainer deviceap = Wifi.Install (WifiPhy, WifiMac, WifiApNode);
\end{lstlisting}  
The first two lines create the default PHY and MAC protocols. Then, line 3 and line 8 create new NetDevices labeled as devicesta and deviceap which are used within the Wi-Fi stations. This code creates the configuration of IEEE 802.11 in infrastructure mode. The network identifier is defined in line 4. Line 5 to 8 attach the station NetDevices with the AP, completing the WLAN configuration.
\subsection{Point-to-Point implementation in NS-3}
\label{Point-to-Point implementation in NS-3}
The NS-3 Point-to-Point (P2P) model is a simplistic model of a point to point serial line link. Data is encapsulated in the Point-to-Point Protocol (PPP - RFC 1661). The key is that the point-to-point protocol link is assumed to be established and authenticated at all times. Simulation of P2P requires that node objects have $ns3::PointToPointNetDevice$ objects installed on them and individual node objects are connected using $ns3::PointToPointChannel$ object. This P2P channel models two wires transmitting bits at the data rate specified by the source net device. 
\begin{figure}[tb]
	\centering
	\includegraphics[width=0.6\textwidth]{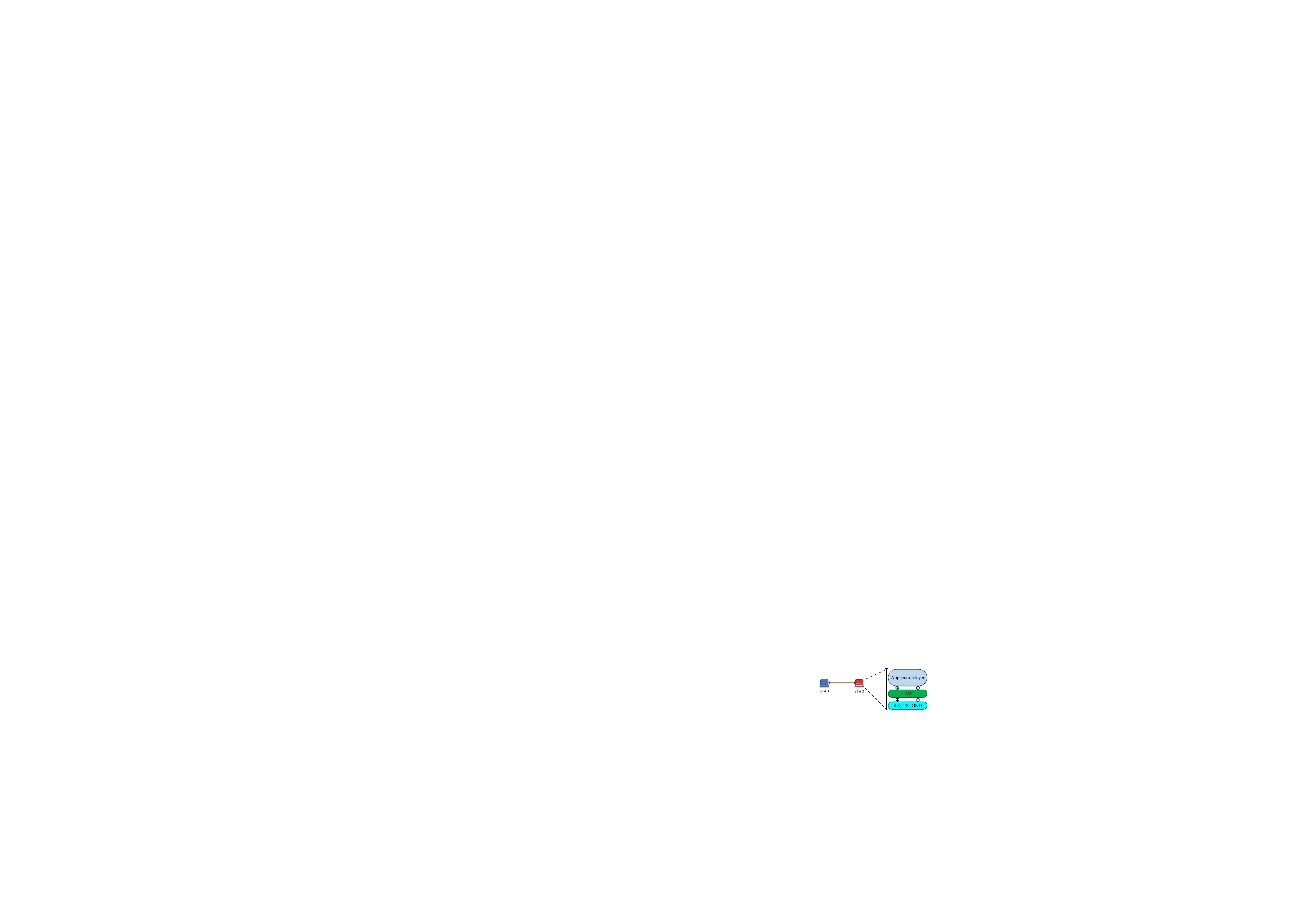}
	\caption{PointToPoint protocol layers.}
	\label{fig4}
\end{figure}
\subsubsection{NS-3 Point-to-Point code}
The following simple code is used to create a P2P link in NS-3:
 \begin{lstlisting}[numbers=left, breaklines=true]
PointToPointHelper p2p;
NetDeviceContainer apeNB = p2p.Install(WifiApNode, eNBNodes)
 \end{lstlisting} 
\subsection{Tracing using Callback functions in NS-3}
The main tracing system provided by NS-3 is callback-based tracing. A callback is an NS-3 object that operates similarly to a function pointer with additional functionality. Each trace source is associated with a specific object class and is identified by a name. A programmer can register a C++ function or method to be called when a certain trace produces a new event. The common uses of callback-based tracing include: 1) Extracting raw data based on events and writing it to a file, 2) Collecting statistics based on certain events and writing them to a file, 3)  Reacting to the event and changing some parameter in the simulation in real time.

In Sections \ref{Methodology to create LWA} and \ref{Methodology to create LWIP}, the detail description of different callback functions used within the design of LWA and LWIP is provided.
\subsection{Packet tags in NS-3}
\label{Packet tags in NS-3}
A tag is a small chunk of information that can be added within a packet generated by the NS-3 simulator without increasing the packet size  (i. e. the duration of packet on the medium is not changed when a tag is included). These tags help in adding simulation information to a packet. The general structure of tags in NS-3 is,
\begin{lstlisting}[numbers=left, breaklines=true]
Ptr <Packet>;
MyTag tag;
p-> AddTag(tag)
p->PeekTag(tag)
\end{lstlisting} 
\chg{In our implementation, tags are only visible to EnB and are used to forward the splitting functionality enabled at the PDCP layer (i.e. LWA is not, is partially or is fully activated) and the RBID of a packet to the LWAAP}.

\subsection{Packet headers in NS-3}
\chg{NS-3 defines packet headers for different protocols used within the simulations. 
The packet classes inherit the NS-3 Header class and provides the interface for other classes to interact with the packets.
Following is the implementation of a new subclass of the ns3::Header base class to represent a protocol header,}
\begin{lstlisting}[numbers=left, breaklines=true]
class NewHeader : public Header
{
public:
  static TypeId GetTypeId (void);
  virtual TypeId GetInstanceTypeId (void) const;
  virtual uint32_t GetSerializedSize (void) const;
  virtual void Serialize (Buffer::Iterator start) const;
  virtual uint32_t Deserialize (Buffer::Iterator start);
  virtual void Print (std::ostream &os) const;
  void SetData (uint32_t data);
  uint32_t GetData (void) const;
  private:
  uint32_t m_data;
};
\end{lstlisting}
The protocol header can attached to a packet by using,
\begin{lstlisting}[numbers=left, breaklines=true]
Ptr<Packet> p = ...;
YHeader yHeader;
yHeader.SetData (0xdeadbeaf);
p->AddHeader (yHeader);
\end{lstlisting}
This header can be extracted from the packet by using the following code,
\begin{lstlisting}[numbers=left, breaklines=true]
Ptr<Packet> p = ...;
YHeader yHeader;
p->RemoveHeader (yHeader);
uint32_t data = yHeader.GetData ();
\end{lstlisting}
\chg{In order to keep a track of the sequence number of packets that are transmitted through LWA or LWIP interface, we use an already existing header for UDP client/server application (called $SeqTsHeader$). To generate the LWAAP PDUs that encompass the status of $lwaactivate$ and RBID, a new header (called $lwaheader$) was defined.} Similarly, an $lwipheader$ is defined that contains $lwipactivate$ and RBID information. This header is inserted by LWIPEP protocol to generate LWIP PDUs.
\subsection{IPSec tunnel implementation in NS-3}
\label{IPSec tunnel implementation in NS-3}
Tunnels are used to create the illusion for some protocols running on a network device that they have a direct connection to another device even when no direct physical link exist between them. In NS-3, the idea of tunneling is achieved through the use of Virtual net devices (called VirtualNetDevice). The main idea is to place an adaptation of interface on top of the real network interfaces. This adaptation allows the use of a unique IP address over different technologies. The virtual net device delegates the task of actual packet transmission to a user callback function. It also allows the user defined code to inject the packets as if they had been received by the VirtualNetDevice. Together, these features are used to build tunnels. For example, by transmitting packets into a UDP socket, an IP-over-UDP-over-IP tunnel can be created. 

The use of VirtualNetDevice allows the entire original IP packet to be encapsulated with a new packet header with different source and destination addresses. One of the drawbacks of using IPSec tunnel is that tunneling increases the bandwidth constraints of the network.
\subsubsection{NS-3 tunnel code}
The NS-3 implementation of a tunnel is represented by the following code example.
 \begin{lstlisting}[numbers=left, breaklines=true]
class Tunnel
{
  Ptr<Socket> m_n1Socket;
  Ptr<Socket> m_n2Socket;
  Ipv4Address m_n1Address;
  Ipv4Address m_n2Address;

  Ptr<VirtualNetDevice> m_n1Tap;
  Ptr<VirtualNetDevice> m_n2Tap;

  bool
  N1VirtualSend (Ptr<Packet> packet, const Address& source, const Address& dest, uint16_t protocolNumber)
  bool
  N2VirtualSend (Ptr<Packet> packet, const Address& source, const Address& dest, uint16_t protocolNumber)
  void N2SocketRecv (Ptr<Socket> socket)
  void N1SocketRecv (Ptr<Socket> socket)
\end{lstlisting} 
The starting and ending point of the tunnel are set as the node containers over which the tunnel is intended to be made. The IP addresses of the netdevice container corresponding to the node container are also passed to the tunnel class. Within the tunnel, VirtualNetDevice is defined and configured to send packets.
\subsection{Flow Monitor in NS-3}
In NS-3, the task for automated result gathering is performed by using a network monitoring feature called Flow Monitor (FlowMonitor). It is integrated in NS-3 as a module and collects/stores network performance data based on per-flow statistics at the IP layer including throughput and latency. etc. It is important to mention here that the main difference between FlowMonitor and PCAP is that the former provide per flow end-to-end statistics, whereas, the later is used to capture intermediate traffic being transported over the medium. Therefore, PCAP packet length is expected to be greater as it contains different technology headers used for transportation.  

The problem with the FlowMonitor is that it is not designed to be used for protocols/mechanism operating in loopback mode. 
\subsubsection{NS-3 FlowMonitor code}
The syntax to define flow probes, to automatically classify
simulated traffic, and to collect flow statistics is shown below.
\begin{lstlisting}[numbers=left, breaklines=true]
FlowMonitorHelper flowmon;
Ptr<FlowMonitor> monitor = flowmon.Install(c);
 
monitor->CheckForLostPackets ();
Ptr<Ipv4FlowClassifier> classifier = DynamicCast<Ipv4FlowClassifier> (flowmon.GetClassifier ());
std::map<FlowId, FlowMonitor::FlowStats> stats = monitor->GetFlowStats ();
for (std::map<FlowId, FlowMonitor::FlowStats>::const_iterator i = stats.begin (); i != stats.end (); ++i)
{
 Ipv4FlowClassifier::FiveTuple t = classifier->FindFlow (i->first);
 std::cout << "Flow " << i->first << " (" << t.sourceAddress << " -> " << t.destinationAddress << ")\n";
 std::cout << " Tx Bytes: " << i->second.txBytes << "\n";
 std::cout << " Rx Bytes: " << i->second.rxBytes << "\n";
 std::cout << " Throughput: " << i->second.rxBytes * 8.0 / (i->second.timeLastRxPacket.GetSeconds() - i-
 >second.timeFirstTxPacket.GetSeconds())/1024 << " Kbps\n";
}
 monitor->SerializeToXmlFile("scratch/newtrial.flowmon", true, true);
\end{lstlisting} 
FlowMonitorHelper takes care of all the details of creating the single classifier, creating one Ipv4FlowProbe per node, and creating the FlowMonitor instance. It installs the monitor in the nodes, set the monitor attributes, and prints the statistics

\section{Implementation of LWA in NS-3}
\label{Customization to create LWA model in NS-3}
\begin{figure}[tb]
	\centering
	\includegraphics[width=1\textwidth]{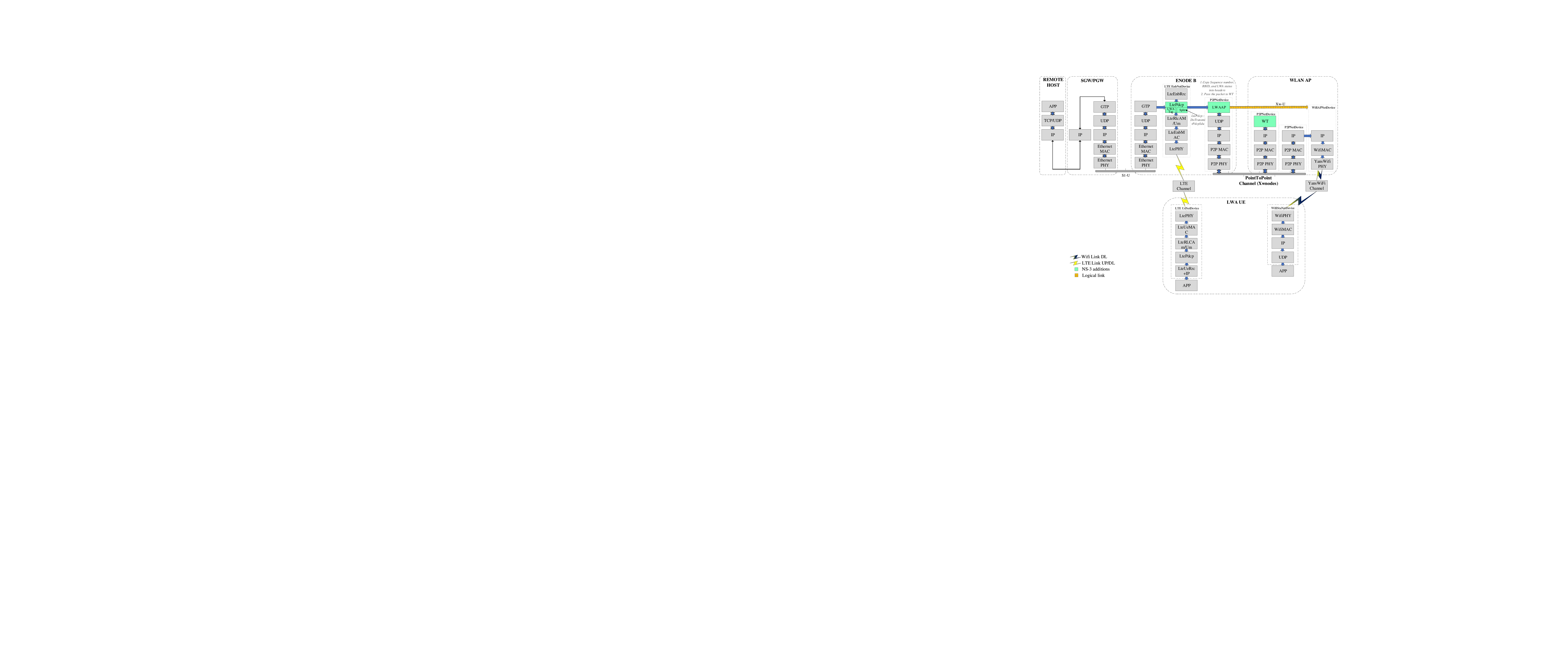}
	\caption{\chg{LWA NS-3 implementation.}}
	\label{fig5}
\end{figure}

As highlighted in Section \ref{Overview of LWA architecture}, the key concept of LWA is that it enables routing packets by using the PDCP layer in the eNB as convergence layer (i.e. a UE in $RRC$-${Connected}$ mode is configured by the eNB to utilize radio resources of LTE and WLAN). The control plane connection remains with the LTE, whereas the data can be routed via both eNB and WLAN. In order to explain how this architecture was simulated in NS-3, we refer to Figure \ref{fig5}, which shows the interconnection between different network devices. Although LWA bearers may be configured to deliver both uplink and downlink data over the WLAN, we focus our study on the implementation of LWA mentioned in release 13 which only focuses on downlink communication. 

Before initiating the simulation, the PDCP layer of LTE eNB can be commanded to activate/deactivate the offloading of data. If LWA is fully activated, the data flow from eNB to UE is stopped and the data is diverted towards the Wi-Fi AP. The aforementioned procedure is accomplished by using the event triggering capabilities of callback functions. 
\subsection{Assumptions}
\label{Assumptions LWA}
In our implementation, we make the following assumptions:
\begin{enumerate}[(1)]
\item The non-collocated LWA scenario is assumed, were the LTE network is connected through a WT to an already existing Wi-Fi network. 
\item We assume an ideal $Xw$-$C$ interface where there is no error in communication and setting up.
\item No mobility model is set for all the the stations (either LTE or WLAN) within the simulation.
\item \chg{
The Wi-Fi AP can accommodate traffic of normal Wi-Fi stations as well as the station used by LWA.}
\item Wifi station is already configured and associated with a WLAN AP.
\end{enumerate}
\subsection{Implementation Details}
\label{Methodology to create LWA}
In this section, we describe the sequence of actions that take place during the simulation of our LWA protocol implementation in NS-3. As shown in Figure \ref{fig5}, multiple network components corresponding to different technologies (LTE, PointToPoint, Wi-Fi) are used together. It also contains various components and features described in Section \ref{Background}. First, any traffic initiated between the remote host and the UE within the EPC model attempts to follow the direct path. However, the downlink traffic can be managed, based on LWA being partially or fully activated. For the case of full LWA being used, all the the traffic is forwarded to the Wi-Fi network. Whereas for partial LWA, some packets are allowed to flow through LTE while others are diverted to flow through Wi-Fi network. 
As shown in Figure \ref{fig5}, two main mechanism representing LWA implementation at the PDCP layer are: LWA Tag and Split. If partial or full LWA is activated, all the packets that are to be passed to Wi-Fi network are tagged to maintain their identity. In the split part, all frames with LWA tags are passed to the P2P link to be forwarded to Wi-Fi network. All the other frames, that are not LWA tagged, are forwarded to the RLC layer.        

\begin{figure}[htb]
	\centering
	\includegraphics[width=1\textwidth]{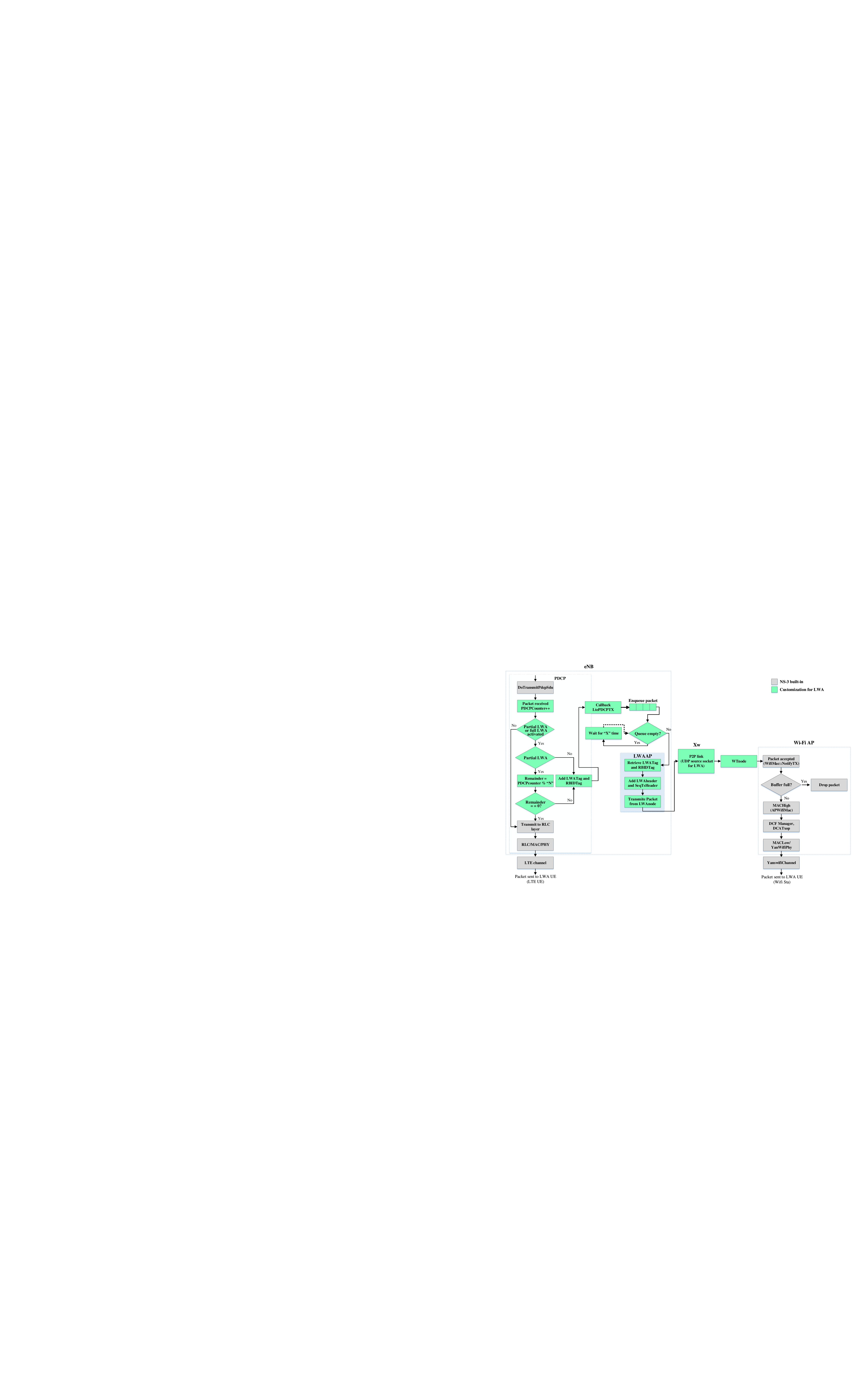}
	\caption{\chg{Activity diagram of LWA implementations in NS-3.}}
	\label{fig5a}
\end{figure}

Figure \ref{fig5} shows a high-level sketch of the implementation, more details of the modifications within the PDCP class along with the design detail of LWA in NS-3 main file are depicted in Figure \ref{fig5a}. This flowgraph provides a detailed description and the subsequent flow of packets of the new additions that connect LTE and Wi-Fi networks in NS-3 (the parts highlighted in green in Figure \ref{fig5} correspond to code additions).
The implementation of the hybrid nature of the LWA architecture is explained as follows; 
\begin{enumerate}[(1)]
\item \textit{Packet generation at the remote host and transfer to PDCP layer}
\label{Packet generation at remote host and transfer to PDCP layer}
The remote host within the EPC model of NS-3 provides IP-based service to a single UE. An OnOff/PacketSink application (that mimics a Voice Over IP (VoIP) traffic flow) is used to transfer as burst of UDP packets over a Client/Server configuration from the remote host to the UE. The Ipv4StaticRoutingHelper class is used to create a static routing table to enable a connection to operate between the remote host and the UE through the eNB. The sequence number of generated packet is attached with the packet as header (called $SeqTsHeader$) within the OnOff packet generator. Following code from $SendPacket ()$ fuction of onoff application shows the process of header attachment.
\begin{lstlisting}[numbers=left, breaklines=true]
 SeqTsHeader seqTs;
 seqTs.SetSeq (m_sent);
 Ptr<Packet> packet = Create<Packet> (m_pktSize); 
 packet->AddHeader (seqTs);
 \end{lstlisting}
 where $m\_sent$ is the counter for sent packets
\item \textit{LWA data flow differentiation}
In NS-3, attributes are used to organize, document and modify the default values used by the various components of the models. In-order to inform the PDCP layer of the eNB about the activation or deactivation of LWA, a new attribute is added. This attribute is called $lwaactivate$. 
  
At the PDCP layer, the new attribute is defined as,  	
\begin{lstlisting}[numbers=left, breaklines=true]
.AddAttribute ("PDCPDecLwa",
 "PDCP LWA decision variable",
 UintegerValue (0),
 MakeUintegerAccessor (&LtePdcp::pdcp_decisionlwa),
 MakeUintegerChecker<uint32_t>())  	
\end{lstlisting}
Add attribute binds the member variable $pdcp$$\_$$decision$ to a public string PDCPDecLwa. It is important to mention here that the value of variable $pdcp$$\_$$decision$ is accessible in the attribute namespace, which is based on string "PDCPDecLwa" and TypeId name ns3::LtePdcp. This variable $pdcp$$\_$$decision$ is used to activate and deactivate the LWA connection. As highlighted in Section \ref{Overview of LWA architecture}, the LWA technique also allows the partial use of both LTE and Wi-Fi technologies together so as to enable in-sequence combining of packets at the UE. The following table enlists different values of $pdcp$$\_$$decision$ and the corresponding stages.
\begin{table}[H]
\centering
\begin{tabular}{|c|c|}
\hline
Value assigned & State                 \\ \hline
0              & LWA deactivated (LTE only)  \\ \hline
1              & Partial LWA (LTE+Wi-Fi) \\ \hline
2              & LWA activated (Wi-Fi only)  \\ \hline
\end{tabular}
\end{table}

These states are also used to to mimic the split and switched bearer operations of LWA (where a $pdcp$$\_$$decision$ value of 1 represents the split bearer and a $pdcp$$\_$$decision$ of 2 indicates the swicthed bearer option). 
At the start of the simulation, the $Config::SetDefault$ method is used to override the initial (default) value of the PDCPDecLwa attribute: 
\begin{lstlisting}[numbers=left, breaklines=true]
Config::SetDefault ("ns3::LtePdcp::PDCPDecLwa", UintegerValue(lwaactivate));
\end{lstlisting}

We currently activate/deactivate the use of LWA before the start of the simulation. However, LWA can be enabled during run-time based on a feedback path from the UE. This procedure would require modifications to be made at the attribute defined within the PDCP layer and can be considered as future work. 
As an alternative, the current approach can also be used to mimic the activation/deactivation based on a feedback. This could be done by running two separate instances of NS-3 LWA simulation in a sequence, so that the outcome of the first could be used as means to enable/disable LWA in the second simulation (while keeping a constant positions of nodes within the two simulations). 

As described in Figure \ref{fig5a}, the first stage of the flowgraph resolves the activation of LWA. If LWA or partial LWA is not activated by setting $pdcp$$\_$$decision$, the received packet is passed to the TransmitPdcpPdu function so that it could be forwarded to the RLC layer. 
\item \textit{RBID and LWA activation status tags}
When $pdcp$$\_$$decision$ variable is assigned a value corresponding to partial or full LWA activate, all packets are tagged with the corresponding RBID and LWA status information before being en-queued for transmission from LWAAP node. Two new tags are created by sub-classing from the abstract base class ns3::Tag (i.e. LWAtag and LCIDtag). This procedure is represented by the entity "Add LWATag and RBIDTag" in Figure \ref{fig5a}. The aforementioned task is achieved by the following code,
\begin{lstlisting}[numbers=left, breaklines=true]
 pdcptag.Set(pdcp_decisionlwa);
 lcidtag.Set((uint32_t)m_lcid);
 p->AddPacketTag (pdcptag);
 p->AddByteTag (lcidtag);
\end{lstlisting}
where $pdcptag$ includes the LWA activation status and $lcidtag$ contains the logical ID of the bearer (RBID is derived from the logical ID).
\item \textit{Flow control in LTE PDCP layer}
Packets upon arriving at the PDCP layer with $pdcp$$\_$$decision$ variable assigned a value corresponding to full LWA activate, are tagged and are en-queued for transmission over the Wi-Fi infrastructure. On the other hand, if the variable $pdcp$$\_$$decision$ is assigned a value corresponding to partial LWA activate, the modulo operator is used to decide the percentage of packets that are to be send to Wi-Fi AP and the eNB\footnote{It is pertinent to mention here that Release 13 of 3GPP does not define the mechanism to split}. This is represented by "Remainder = PDCPcounter \% “N”" in Figure \ref{fig5a}. The variable $N$ can decide the percentage of split. Similar to $lwaactivate$, a new attribute can be defined at the PDCP layer, which can enable $N$ to be set at the start of the simulation. However, in the current implementation, $N$ is set to a fixed value of 2. Thus, when partial LWA is activated, the traffic flow is equally split between LTE and Wi-Fi networks.

In order to allow in-sequence delivery of frames at the LWA UE upper layers, each packet arriving at the PDCP layer of eNB is already assigned a sequence number. This number is embedded within a 12 bytes header called $SeqTsHeader$. The sequence number along with RBID information can be used to aggregate data at the LWA UE.

In order to extract frames from the PDCP layer of eNB, a new callback function is added in the PDCP class of LTE. Given that the same PDCP class is used by the UE, this callback function is only initialized for LTE eNB and is declared as:
\begin{lstlisting}[numbers=left, breaklines=true]
Config::ConnectWithoutContext ("/NodeList/*/DeviceList/*/$ns3::LteeNBNetDevice/LteeNBRrc/UeMap/*/DataRadioBearerMap/*/LtePdcp/TxPDUtrace", MakeCallback (&Lte_pdcpDlTxPDU));
\end{lstlisting}

A new attribute, called TxPDUtrace, is added in the PDCP layer, which gives access to the LWA packet (i.e. Lte\_pdcpDlTxPDU) and is used for the collection of the actual transmission. That is, when ever a packet is tagged with LWAtag and LCIDtag, the trace callback function $LtePdcp::m_pdcptxtrace$ at the PDCP layer is triggered and the packet is passed to the $Lte\_pdcpDlTxPDU$ function. This procedure is represented by "Callback LtePDCPTX" box in Figure \ref{fig5a}.
This packet (with both header and data contents) is copied and en-queued in the buffer for transfer to the Wi-Fi network through a $Xw$ (P2P) link. The new attribute is defined as:
\begin{lstlisting}[numbers=left, breaklines=true]
.AddTraceSource ("TxPDUtrace",
 "PDU to be transmit.",
 MakeTraceSourceAccessor (&LtePdcp::m_pdcptxtrace),
 "ns3::Packet::TracedCallback")
\end{lstlisting}
 
It is important to mention here that the above mentioned callback function is used only for LWA tagged frames and these frames are not passed to the RLC layer of the LTE eNB.
\item \textit{Packet generation to LWAAP}
In 3GPP LTE, a typical Transmit Time Interval (TTI), which refers to the duration of a transmission over the radio link, is set to 1 ms \cite{3GPP}. For a timely and accurate transmission of frames from LTE eNB to Wi-Fi AP, the buffer is continuously observed after every "X" interval. The value of this interval is kept below the aforementioned TTI (i.e. $X$ $= 0.1ms$). Therefore, whenever a packet is placed in the buffer, it is immediately transferred to the source of the UDP socket installed at the LWAAPnode. The aforementioned procedures are represented by "Enqueue packet" "Queue empty" and "Wait for "X" time" entities in Figure \ref{fig5a}. 
\item \textit{Packet formulation by LWAAP and transmission to the $Xw$ link}
As described above, RBID and sequence number of a packets are important parameters that can enable the aggregation of flow at the UE. Information regarding the sequence number is passed along when the packet is copied along with $SeqTsHeader$. RBID and $lwaactivate$ information is first extracted from tags and then added as header within a packet to be transmitted over the $Xw$ interface. 

While tags are used to pass information within a packet in NS-3, they are not part of the packet and do not extend the size of the packet. Thus, in the first stage, LWATag and RBIDTag information is extracted from the packets by using the following code,
\begin{lstlisting}[numbers=left, breaklines=true]
if (m_currentPacket->PeekPacketTag(pdcptag)){
lwastatus = pdcptag.Get();					
}
if(m_currentPacket->FindFirstMatchingByteTag (lcidtag)){  
lcid = lcidtag.Get();
rbid = lcid - 2;	
}
\end{lstlisting}
As described by the 3GPP LWA release 13 standard, the LWAAP entity generates LWAAP PDU containing a RBID identity.
To forward the RBID and LWA status information, a new 2 byte header named $LwaHeader$ is defined. This header is attached to the packet by using the following code,
\begin{lstlisting}[numbers=left, breaklines=true]
LwaHeader lwa;
lwa.SetLwaActivate (lwaactivate);		
lwa.SetBearerId (bearerid);
\end{lstlisting}
In the next stage, the packet is transferred to the source of the UDP socket installed at the LWAAPnode. The above highlighted steps are represented by LWAAP entity in Figure \ref{fig5a}.

\item \textit{Data forward from LWAAPnode to the Wi-Fi station through the Wi-Fi AP}
The downlink flow of a packet from the LWAAPnode to Wi-Fi station is accomplished by creating a source destination socket. The $Xw$ interface in the simulation is assumed to be a P2P link created between the LWAAPnode and the WTnode. Wi-fi AP is also connected to the WT through a P2P link. It is important to mention here that the P2P link of $Xw$-$U$ interface can be replaced with other NS-3 netdevices (such as CSMA) to enable access to outside sockets using tap bridge\footnote{The Tap Bridge is designed to integrate real internet hosts (or more precisely, hosts that support Tun/Tap devices) into NS-3 simulations.} netdevice. This can help in running two separate instance s of NS-3, where one can represent the eNB and LWAAP and the other can include Wi-Fi AP and the Wi-Fi station. Such a setting can be used to reduce the processing strain of running all NS-3 entities within a single machine. 

The LWAAPnode node is set as a source socket and the Wi-Fi station node as the destination sink. Both sockets are identified using IPv4 addresses. The task of the source socket is to connect to the destination socket by identifying it with a given IP address. At the receiver end, the destination socket's task is to bind with the incoming request from the source socket. For synchronization between sockets, the port number for source and destination is kept the same. In order to create the routing database and initialize the routing tables for connection between the P2P node, Wi-Fi AP and Wi-Fi station, the Ipv4GlobalRoutingHelper class was used. 
\item \textit{Packet reception at Wi-Fi station}
The packet received at the destination socket at the Wi-Fi station include the $SeqTsHeader$ and the $LwaHeader$ headers. The $lwaactivate$ status, RBID information (i.e. bearer ID) and sequence number of the received packet can be extracted by using the following code.
\begin{lstlisting}[numbers=left, breaklines=true]
m_currentPacket->RemoveHeader (lwa);
lwaactivate=lwa.GetLwaActivate();
bearerid=lwa.GetBearerId();
m_currentPacket->RemoveHeader (seqTsx); 
currentSequenceNumber = seqTsx.GetSeq ();
\end{lstlisting}
This information can be used to aggregate packets at the application layer of the UE. 
\end{enumerate}
\chg{With the help of Table~\ref{tab:2}, a comparison between implemented functions of LWA in NS-3 with 3GPP LWA standard is provided. As highlighted in the table, most of the specifications of LWA provided by 3GPP release 13 were implemented in NS-3. LWA was controlled at the eNB and the splitting of frames was performed at the PDCP layer. The $pdcp\_decisionlwa$ variable enabled the NS-3 implementation to support split and switched bearer functionality. Since the implementation did not include a method to aggregate data at the PDCP layer, no changes were made at the NS-3 LTE UE\footnote{The aggregation of flows at the PDCP layer required modifications to the core of the NS-3 LTE module and was considered out of scope of our current work.}}. As mentioned in Section \ref{Assumptions LWA}, it was assumed that an ideal control plane ($Xw$-$C$) was available between eNB and WT. Therefore, the details of WLAN measurements, support for Establishment, modification and error handling mechanism were not considered within this implementation. 
\begin{table}[H]
	\caption{\chg{NS-3 LWA implementation compared with 3GPP standard.}}
	\label{tab:2}
	\centering
	{
    	\scalebox{0.9}{
		\begin{tabular}{|>{\centering\arraybackslash}p{6cm}|>{\centering\arraybackslash}p{3cm}|>{\centering\arraybackslash}p{3cm}|}
			\hline
			\parbox[c][5ex]{5ex}{\centering}\bf{Attributes}&\bf{NS-3 LWA}&\bf{3GPP standard LWA}\\ \hline
			\parbox[c][4ex]{4ex}{\centering}eNB control&Yes&Yes  \\\hline       
			\parbox[c][4ex]{4ex}{\centering}Connecting layers&PDCP&PDCP \\\hline
			\parbox[c][4ex]{4ex}{\centering}Offloading granularity&Split or Switched Bearer & Split or Switched Bearer\\\hline
			\parbox[c][4ex]{4ex}{\centering}Upgrade in LTE network&eNB and UE&eNB\\\hline  
			\parbox[c][4ex]{4ex}{\centering}Aggregating flows at UE PDCP&No&Yes\\\hline 
            \parbox[c][4ex]{4ex}{\centering}New network entities in LTE&LWAAP and $Xw$-$U$&LWAAP and $Xw$\\\hline
            \parbox[c][4ex]{4ex}{\centering}Additional interface for flow control&$Xw$-$U$&$Xw$-$U$\\\hline
            \parbox[c][4ex]{4ex}{\centering}$Xw$ control plane interface&No&$Xw$-$C$\\\hline
            \parbox[c][4ex]{4ex}{\centering}
            WT connection establishment&No&Yes\\\hline
            \parbox[c][4ex]{4ex}{\centering}New network nodes in Wi-Fi&WT&WT\\\hline
            \parbox[c][4ex]{4ex}{\centering}WLAN measurements&No&Yes\\\hline
            \parbox[c][4ex]{4ex}{\centering}WLAN security&No&WLAN native 802.1x EAP/AKA with fast authentication method\\\hline
            \parbox[c][4ex]{4ex}{\centering}Regular Wi-Fi services on the same hardware&Yes&Yes\\\hline
            \parbox[c][4ex]{4ex}{\centering}WLAN traffic direction&Downlink&Downlink\\\hline
		\end{tabular}
	}}
\end{table}
\section{Customization to create LWIP model in NS-3}
\label{Customization to create LWIP model in NS-3}
In LWIP, an IPSec tunnel is used to transmit downlink traffic from eNB to the UE through a Wi-Fi AP. Similar to LWA, the UE in LWIP has a RRC connection to the eNB. However, the main difference between LWIP and LWA is that LWIP supports downlink switching of IP flows at the IP layer while LWA is able to aggregate flows at the PDCP layer. Unlike LWA, packets in LWIP can not be simultaneously transmitted to both LTE or WLAN link. As described in Section \ref{LTE implementation in NS-3}, the IP networking for end-to-end connectivity for end users (that contains UE, SGW/PGW and remote host) does not involve the use of IP stack at the eNB in the EPC model of the NS-3. In order to counter the aforementioned problem and to gain access to PDCP SDUs (IP packets), 
the LWIP protocol was designed to follow the same methodology used in Section \ref{Methodology to create LWA} for LWA to extract frames at the PDCP layer.  
\begin{figure}[tb]
	\centering
	\includegraphics[width=0.9\textwidth]{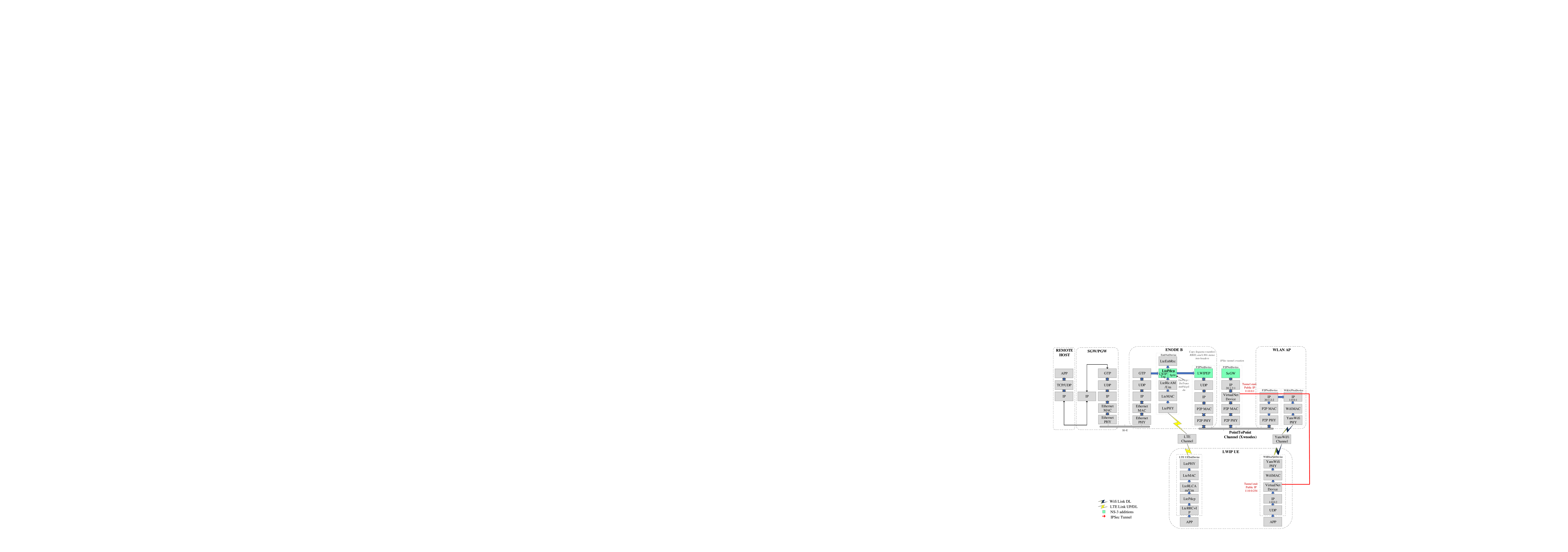}
	\caption{LWIP NS-3 implementation.}
	\label{fig8}
\end{figure}
\subsection{Main Idea}
Similar to LWA implementation, the PDCP layer of LTE eNB can be commanded to activate/deactivate the offloading of data. If LWIP is activated, the complete flow from LTE to UE is stopped and the data is diverted towards the Wi-Fi AP. The aforementioned procedure is accomplished by event driven triggering technique. An IPSec tunnel is created on which IP packets are tunneled over UDP/IP. It is important to mention here that LWIP specifications mentioned in 3GPP Release 13 do not allow the use of split bearer technique. Also, the use of $Xw$ interface is not mandatory. For the NS-3 implementation, since the same LWA mechanism is leveraged, $Xw$ interface was used. This is also justified by the fact that the implemented $Xw$-$U$ interface was only responsible for transfer of packets. 
\subsection{Assumptions}
\label{Assumptions for LWIP}
For our implementation, we make the following assumptions:
\begin{enumerate}[(1)]
\item The non-collocated scenario is assumed, where LTE network is connected through SeGW to an already existing Wi-Fi network. 
\item We assume an ideal $Xw$-$C$ interface where there is no error in communication and setting up.
\item No mobility model is set for all the the stations (either LTE or WLAN) within the simulation.
\item Wi-Fi station is already configured and associated with a WLAN AP.
\item The LWIP tunnel between the LWIP SeGW and the Wi-Fi station is already established through Wi-Fi infrastructure.
\item The IPSec tunnel is an IP tunnel that only contains authentication but does not use encryption. While the packets sent over the tunnel are authenticated by the LWIP tag, adding end-to-end encryption of frames is left as a future work.
\end{enumerate}
\subsection{Methodology to create LWIP}
\label{Methodology to create LWIP}
In this section, we describe the sequence of events that take place during the simulation of LWIP protocol in NS-3. Figure \ref{fig8} indicates that multiple network components corresponding to different technologies (LTE, PointToPoint, VirtualNetDevice and Wi-Fi) are used together to develop the LWIP protocol. An IP tunnels is created using a VirtualNetDevice that wraps UDP packets with new IP headers. The significance of different technologies used together in Figure \ref{fig8} is explained as follows.

First, any traffic generated from the remote host to the UE attempts to follow the direct path. However, if the downlink traffic is to be forwarded to the Wi-Fi network, a secure tunnel is used to funnel the data to the LWIP UE (i.e. the Wi-Fi station).

\begin{figure}[htb]
	\centering
	\includegraphics[width=1\textwidth]{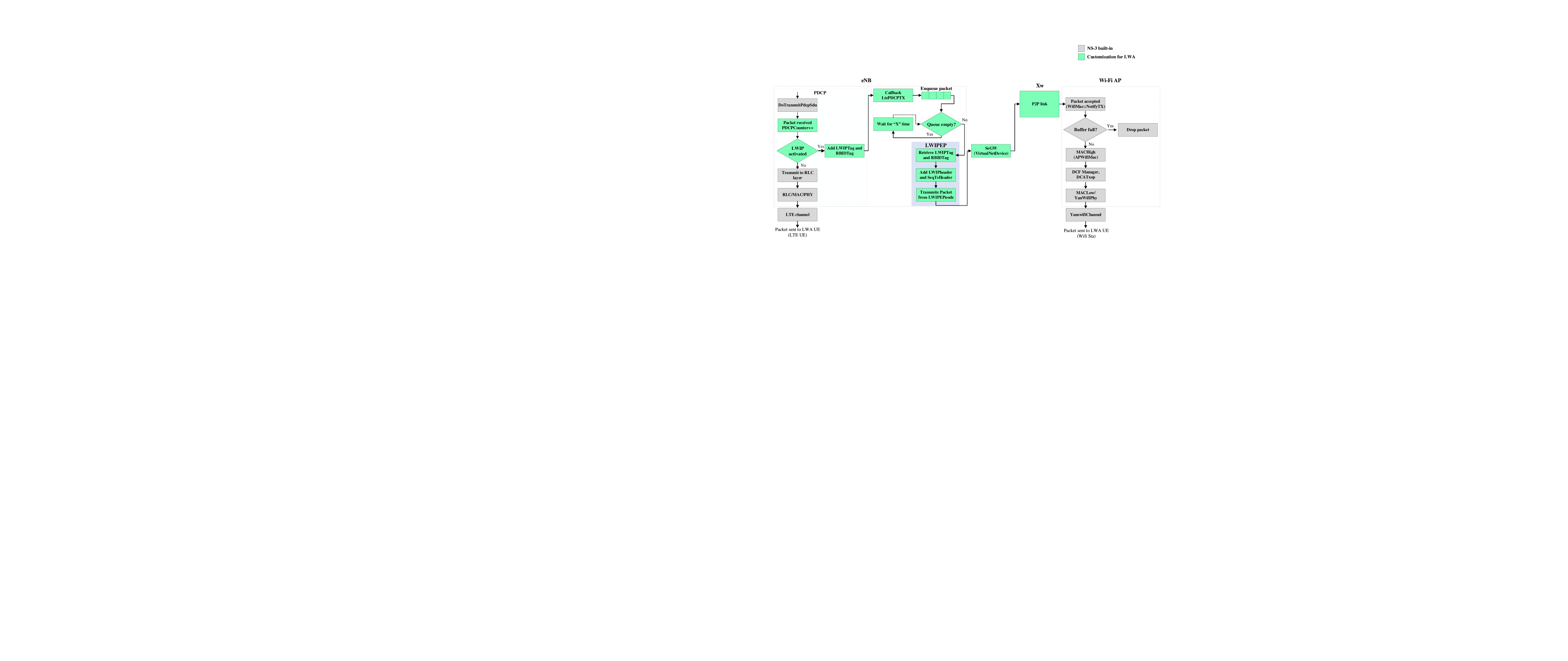}
	\caption{Activity diagram of LWIP implementations in NS-3.}
	\label{fig9a}
\end{figure}

The details of modifications within PDCP classes and the interconnection between different classes is highlighted in Figure \ref{fig9a}. 
This flowgraph provides a detailed description and the subsequent flow of packets over the new network nodes/modules added to connect LTE and Wi-Fi networks through the IP tunnel (the parts highlighted as NS-3 additions in Figure \ref{fig8}).
The implementation of the hybrid nature of the LWIP architecture is explained as follows; 
\begin{enumerate}[(1)]
\item \textit{Packet generation at remote host and transfer to PDCP layer}
An OnOff/PacketSink application (that mimics a Voice Over IP traffic flow) is used to transfer a burst of UDP packets over a Client/Server configuration from the remote host to the LTE UE. The Ipv4StaticRoutingHelper class is used to create a static routing table to create a connection from the remote host to the UE through eNB. The sequence number of generated packet is attached with the packet as header. As mentioned above, $SeqTsHeader$ is the header that contains the aforementioned information. 
\item \textit{LWIP data flow differentiation}
In order to inform the PDCP layer of the eNB about the activation or deactivation of LWIP, a new attribute is added. This attribute is called lwipactivate.

At the PDCP layer, the new attribute is defined as,  	
\begin{lstlisting}[numbers=left, breaklines=true]
.AddAttribute ("PDCPDecLwip",
 "PDCP LWIP decision variable",
 UintegerValue (0),
 MakeUintegerAccessor (&LtePdcp::pdcp_decisionlwip),
 MakeUintegerChecker<uint32_t>())  	
  \end{lstlisting}
This variable $pdcp$$\_$$decision$ is used to activate and deactivate the LWIP connection. The following table enlists different values of $pdcp$$\_$$decision$ and the corresponding  stages.
\begin{table}[H]
\centering
\begin{tabular}{|c|c|}
\hline
Value assigned & State                 \\ \hline
0              & LWIP deactivated (LTE only)  \\ \hline
1              & LWIP activated (Wi-Fi only)  \\ \hline
\end{tabular}
\end{table}
At the start of the simulation, the Config::SetDefault method is used to override the initial (default) value of the PDCPDecLwip attribute. This change in value is achieved by the following code,   
\begin{lstlisting}[numbers=left, breaklines=true]
Config::SetDefault ("ns3::LtePdcp::PDCPDecLwip", UintegerValue(lwipactivate));
\end{lstlisting}
As described in Figure \ref{fig9a}, the first stage of the flowgraph is used to take a decision about the activation of LWIP. If LWIP is not activated by setting $pdcp$$\_$$decision$, the received packet is passed to the $TransmitPdcpPdu$ function so that it could be forwarded to the RLC layer. 
\item\textit{RBID  and  LWA  activation  status  tags}
Similar to LWA, when $pdcp\_decision$ variable is assigned a value corresponding to LWIP activate, all packets are tagged with the corresponding RBID and LWIP status information before being en-queued for  transmission  from  LWIPEP node. The LCIDtag defined for LWA is reused. However, for LWIP status, a new tag (called LWIPTag) is defined. This procedure is represented by the entity ”Add LWIPTag and RBIDTag” in Figure 10.  
\item \textit{Flow control in LTE PDCP layer}
Packets upon arriving at the PDCP layer with $pdcp$$\_$$decision$ variable assigned a value corresponding to LWIP activate, are tagged and are en-queued for transmission over the Wi-Fi Infrastructure. 

In  order  to  allow  in-sequence  delivery  of  frames  at  the  LWA  UE  upper layers, each packet arriving at the PDCP layer of eNB is already assigned a  sequence  number.   This  number  is  embedded  within  a  12  bytes  SeqTsHeader.

In order to extract the LWIP tagged frames from the PDCP layer, the same approach followed in step 4 of Section \ref{Methodology to create LWA} was used. 
\item \textit{Packet generation to LWIPEP entity}
Step 5 of Section \ref{Methodology to create LWA} was reused to immediately transferred available packets to the LWIPEPnode. 
\item \textit{Packet  formulation  by  LWIPEP  and  transmission  to  SeGW  link}
As mentioned in Section \ref{Overview of LWIP architecture}, the responsibility of LWIPEP module is to generate user plane data along with the bearer identification. Similar to LWA mechanism, RBID and lwip activate information is first extracted from tags and then added as header with the packet to be transmitted over the $Xw$ interface. A new 2 byte header named $LwipHeader$ is defined (similar to $LWAHeader$) that contains lwip activate and bearer ID information. This stage is represented by LWIPEP entity in Figure \ref{fig9a}. 

In the next phase, the packet is transferred to the SeGW from LWIPEPnode for transmission over the IPSec tunnel.
\item \textit{Data forward from SeGW  to Wi-Fi station through IP tunnel}
To create an IP tunnel, extra interfaces (virtual interfaces) were installed over the SeGW link and the Wi-Fi station (i.e. an interface with a new address set to 11.0.0.1 was installed over the P2P link and the interface with address 11.0.0.254 was installed over the Wi-Fi station). Thus, the flow of packets would be between 11.0.0.x (i.e. the tunnel) instead of the actual IP addresses (10.0.x.y for P2P and 192.168.x.y for the Wi-Fi network). The use of VirtualNetDevice at the SeGW represents a contact point towards external Wi-Fi network. This is the point where the secured IPSec tunnel is initiated. A tunnel class is created that utilizes callback functions to create virtual UDP source and destination sockets over the VirtualNetDevice. These sockets are used to transmit the IP encapsulated LWIP from SeGW to Wi-Fi station. 

The $Xw$ interface defined for LWA is used to provide a connection between SeGW and Wi-Fi AP.
\item \textit{Packet reception at Wi-Fi station}
The packet received at the destinatio at the Wi-Fi station include the $SeqTsHeader$ and the $LwipHeader$ headers. The $lwaactivate$ status, RBID information (i.e. bearer ID) and sequence number of the received packet can be extracted by using the following code.
\begin{lstlisting}[numbers=left, breaklines=true]
m_currentPacket->RemoveHeader (lwip);
lwipactivate=lwip.GetLwaActivate();
bearerid=lwip.GetBearerId();
m_currentPacket->RemoveHeader (seqTsx); 
currentSequenceNumber = seqTsx.GetSeq ();
\end{lstlisting}
This information can be used to aggregate packets at the application layer of the UE.
\end{enumerate}

With the help of Table~\ref{tab:3}, a comparison between implemented functions of LWA in Ns-3 with 3GPP standard is provided. As highlighted in the table, most of the specifications of LWIP provided by 3GPP release 13 were implemented in NS-3. Activation and deactivation of LWIP mechanism was controlled by the eNB. Due to the problem of non-existing IP stack at the eNB, the splitting of IP packets was performed at the  PDCP layer. The $pdcp\_decisionlwip$ variable enabled the NS-3 implementation to support switched bearer functionality. Since the implementation did not include a method to aggregate data at the IP layer, no changes were made at the NS-3 LTE UE. As mentioned in Section \ref{Assumptions for LWIP}, it was assumed that an ideal control plane ($Xw$-$C$) was available between SeGW and Wi-Fi AP. Therefore, the details of WLAN measurements, support for establishment, modification and error handling mechanism were not considered within this implementation. Although LWIP can be configured to deliver both uplink and downlink data over the WLAN, we focused our study on the implementation particularly for downlink communication.
\begin{table}[tb]
	\caption{\chg{NS-3 LWIP implementation compared with 3GPP standard.}}
	\label{tab:3}
	\centering
	{
    	\scalebox{0.9}{
		\begin{tabular}{|>{\centering\arraybackslash}p{6cm}|>{\centering\arraybackslash}p{3cm}|>{\centering\arraybackslash}p{3cm}|}
			\hline
			\parbox[c][5ex]{5ex}{\centering}\bf{Attributes}&\bf{NS-3 LWIP}&\bf{3GPP standard LWIP}\\ \hline
			\parbox[c][4ex]{4ex}{\centering}eNB control&Yes&Yes  \\\hline       
			\parbox[c][4ex]{4ex}{\centering}Connecting layers&PDCP&IP (PDCP SDU) \\\hline
			\parbox[c][4ex]{4ex}{\centering}Offloading granularity&Switched Bearer &Switched Bearer\\\hline
			\parbox[c][4ex]{4ex}{\centering}Upgrade in LTE network&eNB&eNB\\\hline  
			\parbox[c][4ex]{4ex}{\centering}Aggregating flows at UE IP&No&Yes\\\hline 
            \parbox[c][4ex]{4ex}{\centering}New network entities in LTE&LWIPEP and SeGW&LWIPEP and SeGW\\\hline
            \parbox[c][4ex]{4ex}{\centering}New network nodes in Wi-Fi&None&None\\\hline
            \parbox[c][4ex]{4ex}{\centering}WLAN measurements&No&Yes\\\hline
            \parbox[c][4ex]{4ex}{\centering}WLAN security&No&WLAN native 802.1x EAP/AKA with fast authentication method\\\hline
            \parbox[c][4ex]{4ex}{\centering}Regular Wi-Fi services on the same hardware&Yes&Yes\\\hline
            \parbox[c][4ex]{4ex}{\centering}WLAN traffic direction&Downlink&Downlink plus uplink\\\hline
		\end{tabular}
	}}
\end{table}
\section{Results of LWA and LWIP for single instance of NS-3 simulation}\label{results}
\begin{figure}[H]
	\centering
	\includegraphics[width=.4\textwidth]{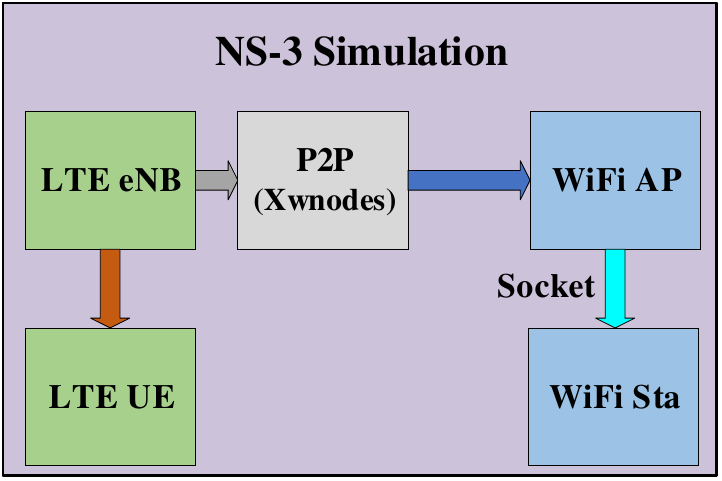}
	\caption{Single instance of NS-3.}
	\label{sinin}
\end{figure}

 In this section, we describe some of the results for the LWA and LWIP implementation for the case when all entities reside in a single instance of NS-3 simulation. Figure \ref{sinin} describes the aforementioned scenario.
\begin{figure}[H]
	\centering
	\includegraphics[width=0.8\textwidth]{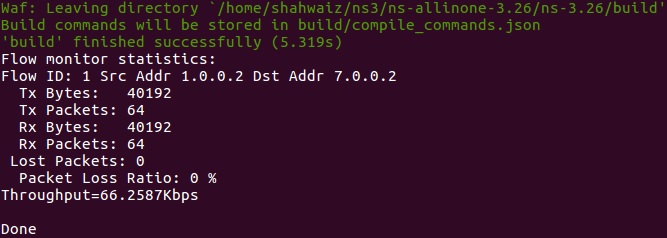}
	\caption{LWA and LWIP deactivated.}
	\label{fig10}
\end{figure}
A single VoIP application with data rate of 64 Kbps and packet size of 600 bytes was initiated from the remote host to the UE. The application is run in saturation conditions for 4.825 seconds. We start by first showing the case when both LWA and LWIP techniques are deactivated.  

Figure \ref{fig10} shows the FlowMonitor output of the aforementioned case. The packet length received with IP and UDP header is 628 bytes (i.e. 8 Bytes for UDP header and 20 Bytes for IP header).
\subsection{LWA NS-3 Implementation}
\begin{figure}[H]
	\centering
	\includegraphics[width=0.8\textwidth]{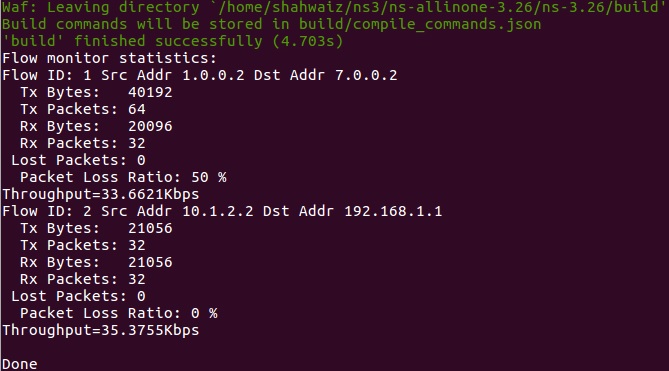}
	\caption{Partial LWA activated.}
	\label{fig11}
\end{figure}

Figure \ref{fig11} shows the flow monitor results when half of the VoIP flow is diverted to the Wi-Fi network and the other half follows directly to the UE. 
\begin{figure}[H]
	\centering
	\includegraphics[width=1\textwidth]{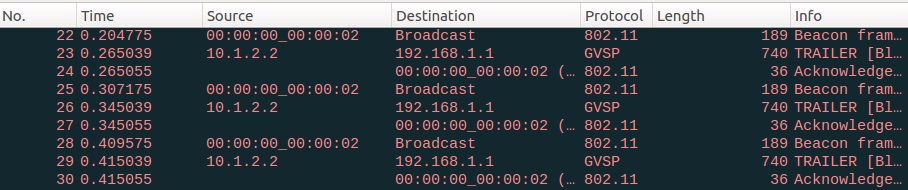}
	\caption{Glimpse of PCAP for LWA scheme.}
	\label{fig12}
\end{figure}
Even though half of the packets are transferred to the Wi-Fi network, the packet length over the Wi-Fi network increases to 658 bytes. This is because packets are copied at the PDCP layer which contains 30 bytes of PDCP and RRC header. Figure \ref{fig12} shows the PCAP over the Wi-Fi channel. Packet length, in this case is 740, which includes the 82 bytes of Wi-Fi headers and 28 bytes of the IP/UDP header.
\subsection{LWIP NS-3 Implementation}
\begin{figure}[tb]
	\centering
	\includegraphics[width=0.8\textwidth]{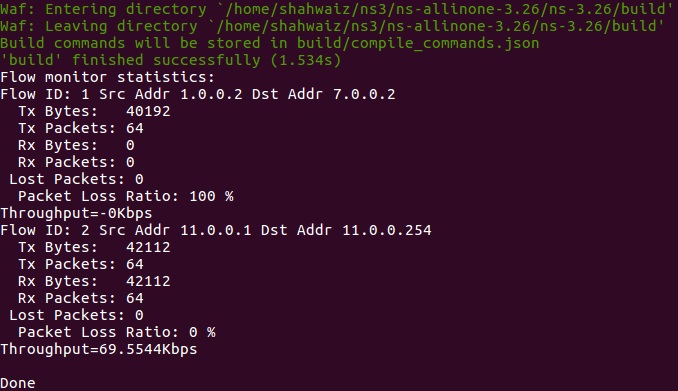}
	\caption{LWIP activated.}
	\label{fig13}
\end{figure}

\begin{figure}[htb]
	\centering
	\includegraphics[width=1\textwidth]{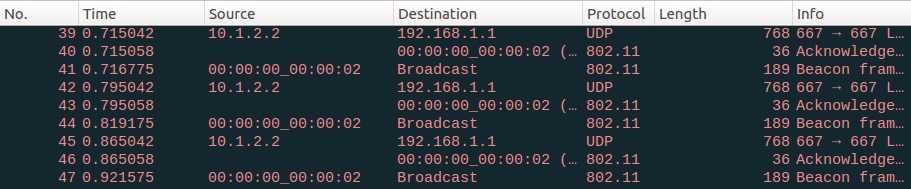}
	\caption{Glimpse of PCAP for LWIP scheme.}
	\label{fig14}
\end{figure}

For LWIP implementation, we consider the complete transfer of packets from LTE to the Wi-Fi network (no traffic splitting is considered as defined by the 3GPP). Figure \ref{fig13} indicates the flow monitor results for the aforementioned case. These results correspond to the packets received at the IP layer of the LWIP Wi-Fi station. Thus, the packet size is similar to LWA flow monitor results (i.e. 658 bytes) and it only contains the additional UDP/IP header. As shown in Figure \ref{fig8}, the VirtualNetDevice in installed on Wi-Fi station below the UDP and IP layer, the packet size does not include the additions due to virtual net devices.
However, the PCAP results shown in Figure \ref{fig14} indicate an increase of 28 bytes when compared with LWA PCAP. This corresponds to the additional UDP and IP header used to create the tunnel. 
\section{Results of LWA for two instance of NS-3 simulation}
\label{Results of LWA and LWIP for two instance of NS-3 simulation}
\begin{figure}[H]
	\centering
	\includegraphics[width=.4\textwidth]{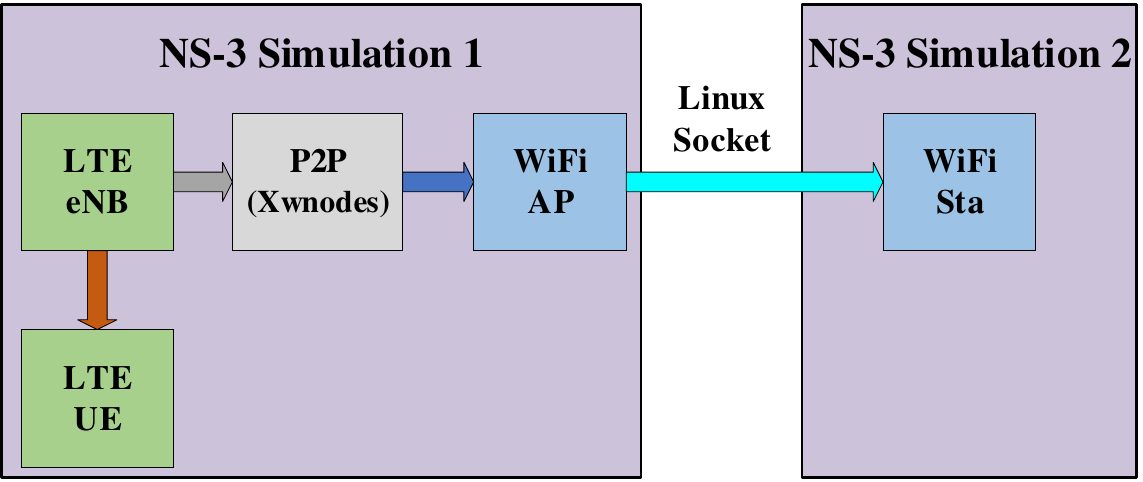}
	\caption{Two instance of NS-3.}
	\label{twoin}
\end{figure}

As described in \ref{Methodology to create LWA}, the downlink flow of packets from LWAAPnode to Wi-Fi station is achieved with the creation of source destination socket. This source and destination sockets could also be replaced with Linux socket, so as to enable to separate NS-3 instances to communicate with each other. Figure \ref{twoin} describes the details of the aforementioned scenario. Lte EPC model along with Wi-Fi AP and the modification to enable LWA and LWIP are placed in first instance of NS-3 and the second instance of NS-3 contains the Wi-Fi station. The source socket is created in Xwnodes and destination socket is placed at Wi-Fi station. 
\subsection{Partial LWA results}
\begin{figure}[H]
	\centering
	\includegraphics[width=0.8\textwidth]{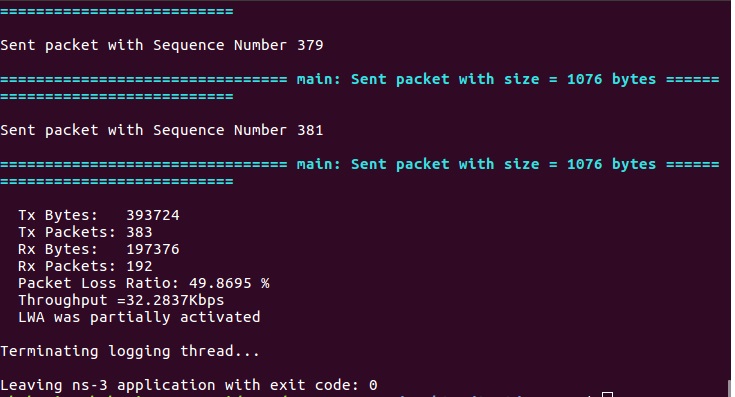}
	\caption{Results for first instance of NS-3 for partial LWA.}
	\label{partenbap}
\end{figure}

In the section, we present results for th case when LWA is partially activated. Figure \ref{partenbap} represents the results for first instance of NS-3 and indicates that half of the packet are transmitted to the LTE UE.
\begin{figure}[H]
	\centering
	\includegraphics[width=0.8\textwidth]{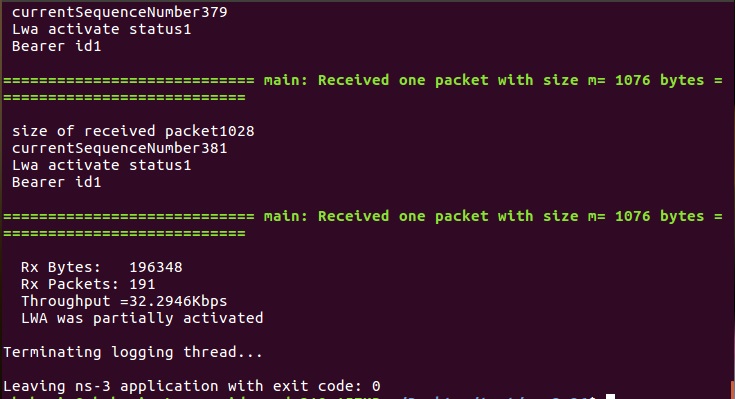}
	\caption{Results for second instance of NS-3 for partial LWA.}
	\label{partensta}
\end{figure}
Figure \ref{partensta} shows that half of the packets are passed through LWAAPnode to Wi-Fi stations. It is pertinent to highlight here that the $SeqTsHeader$ header is used to extract the sequence number of packet received by the Wi-Fi station. Also, $LwaHeader$ header is used to gain information regarding bearer id and status of LWA procedure. 
\section{Conclusions}\label{conclusions}
In this article, we  present the design details of our implementation of LTE-WLAN offloading techniques (i.e. LWA and LWIP) proposed in release 13 by 3GPP. We provide details of the two techniques and describe the building blocks used to implement them in NS-3. We provide a step-by-step description of different aspects of the implementation. Through results, we showcase the viability of the implementation. The main difference between the LWA and LWIP implementation are as follows: 1) In LWA technique, flow can be simultaneously transmitted in both the LTE or WLAN link. This feature can allow a UE to enjoy the aggregated bandwidth of LTE and Wi-Fi. However, the in-sequence delivery of packets requires extra complexity at both the eNB and the UE, where packets are tagged for each path. 2) While LWIP provides secure means of data delivery and provides universal solution which can work with any Wi-Fi node, it increases the bandwidth constraints of the network. Also, complete use of either LTE or Wi-Fi path makes this technique less flexible to be used by the UE. 
\bibliographystyle{abbrv}

\end{document}